\documentclass[aps,prl,twocolumn,superscriptaddress]{revtex4-1}
\usepackage{times}
\usepackage{comment}
\usepackage{graphicx}
\usepackage{feynmf}
\usepackage{tabularx}
\usepackage{amsmath}
\usepackage{amstext}
\usepackage{amssymb}
\usepackage{xfrac}
\usepackage[colorlinks,citecolor=blue]{hyperref}
\usepackage{graphicx}
\usepackage{amsmath}
\usepackage{amstext}
\usepackage{amssymb}
\usepackage{amsfonts}
\usepackage{longtable,booktabs}
\usepackage{hyperref}
\usepackage{url}
\usepackage{subfigure}
\usepackage{dsfont}
\usepackage{booktabs}
\usepackage{amsbsy}
\usepackage{dcolumn}
\usepackage{amsthm}
\usepackage{bm}
\usepackage{esint}
\usepackage{multirow}
\usepackage{hyperref}
\usepackage{cleveref}
\usepackage{mathrsfs}
\usepackage{amsfonts}
\usepackage{amsbsy}
\usepackage{dcolumn}
\usepackage{bm}
\usepackage{multirow}
\usepackage{color}
\usepackage{extarrows}
\usepackage{datetime}
\usepackage[super]{nth}
\hypersetup{
	colorlinks=magenta,
	linkcolor=blue,
	filecolor=magenta,
	urlcolor=magenta,
}

\begin{document}
\title{Yang-Lee Zeros, Semicircle Theorem, and Nonunitary Criticality in Bardeen-Cooper-Schrieffer Superconductivity}
\author{Hongchao Li}
\thanks{These two authors contributed equally to this work.}
\affiliation{Department of Physics, University of Tokyo, 7-3-1 Hongo, Tokyo 113-0033,
Japan}
\email{lhc@cat.phys.s.u-tokyo.ac.jp}

\author{Xie-Hang Yu}
\thanks{These two authors contributed equally to this work.}
\affiliation{Max-Planck-Institut für Quantenoptik, Hans-Kopfermann-Straße 1, D-85748
Garching, Germany}
\affiliation{Munich Center for Quantum Science and Technology, Schellingstraße
4, 80799 München, Germany}
\email{xiehang.yu@mpq.mpg.de}

\author{Masaya Nakagawa}
\affiliation{Department of Physics, University of Tokyo, 7-3-1 Hongo, Tokyo 113-0033,
Japan}
\email{nakagawa@cat.phys.s.u-tokyo.ac.jp}

\author{Masahito Ueda}
\affiliation{Department of Physics, University of Tokyo, 7-3-1 Hongo, Tokyo 113-0033,
Japan}
\affiliation{RIKEN Center for Emergent Matter Science (CEMS), Wako, Saitama 351-0198,
Japan}
\affiliation{Institute for Physics of Intelligence, University of Tokyo, 7-3-1
Hongo, Tokyo 113-0033, Japan}
\email{ueda@cat.phys.s.u-tokyo.ac.jp}

\date{\today}
\begin{abstract}
Yang and Lee investigated phase transitions in terms of zeros of partition functions, namely, Yang-Lee zeros~[\href{https://journals.aps.org/pr/abstract/10.1103/PhysRev.87.404}{Phys. Rev. \textbf{87}, 404 (1952)}; \href{https://journals.aps.org/pr/abstract/10.1103/PhysRev.87.410}{ Phys. Rev. \textbf{87}, 410 (1952)}]. We show that  the essential singularity in the superconducting gap is directly related to the number of roots of the partition function of a BCS superconductor. Those zeros are found to be 
distributed on a semicircle in the complex plane of the interaction strength due to the Fermi-surface instability. A renormalization-group analysis shows that the semicircle theorem holds for a generic quantum many-body system with a marginal coupling, in sharp contrast
with the Lee-Yang circle theorem for the Ising spin system. This indicates that the geometry of Yang-Lee zeros is directly connected to the Fermi-surface instability. Furthermore, we unveil the nonunitary criticality in BCS superconductivity that emerges at each individual Yang-Lee zero due to exceptional points and presents a universality class distinct from that of the conventional Yang-Lee edge singularity. 
\end{abstract}
\maketitle
\emph{Introduction}.---
Yang and Lee developed a general approach to understanding phase transitions in terms of zeros, known as Yang-Lee zeros, of the partition function \cite{PhysRev.87.404,PhysRev.87.410}. 
They investigated the distribution of zeros of the partition function of a classical Ising model for
an imaginary magnetic field to understand the mathematical origin of nonanalyticity of the ferromagnetic phase transition. 
The thermal phase transition between the paramagnetic and ferromagnetic phases occurs when the distribution of zeros 
touches the real axis in the thermodynamic limit. Yang-Lee zeros are also closely related to singularities in thermodynamic quantities accompanied by anomalous scaling \cite{Fisher:1978vn,Kurtze:1979wb,10.1143/PTP.69.14,Cardy:1985ub,Cardy:1989uo,Zamolodchikov:1991tl}.
This type of singularities in critical phenomena is collectively referred to as the Yang-Lee singularity \cite{BENA2005}.

The distribution of Yang-Lee zeros governs the critical phenomena in phase transitions \cite{Fisher1965,Fisher:1978vn} and is of fundamental importance in statistical physics. The universality of the distribution is encapsulated by the Lee-Yang circle theorem \cite{PhysRev.87.404,PhysRev.87.410}, which states that the Yang-Lee zeros of the ferromagnetic Ising model are distributed on a unit circle in the complex plane of the fugacity \cite{Simon:1973tr,Newman:1974wi,Lieb:1981vb,Kortman:1971tw}. While the Yang-Lee theory has been applied to a wide range of phase transitions in classical \cite{PhysRevLett.84.4794,PhysRevLett.84.814,PhysRevLett.89.080601,PhysRevLett.110.248101} and quantum \cite{Gehlen_1991,Sumaryada:2007uu,PhysRevB.53.7704,Matsumoto2020,PhysRevResearch.3.033206,PhysRevB.106.054402,PhysRevE.96.032116,PhysRevX.11.041018,PhysRevE.96.032116,Fredrik2023,PhysRevLett.131.080403} systems, its application to itinerant electronic systems is limited \cite{Sumaryada:2007uu,PhysRevB.53.7704}. Itinerant electrons show various types of order arising from Fermi-surface instabilities, including the Bardeen-Cooper-Schrieffer (BCS) superconductivity \cite{Bardeen:1957tx} as a prime example. Thus, the study of Yang-Lee zeros in the BCS theory is expected to unveil hitherto unnoticed universality in superconductivity.

In this Letter, we develop the Yang-Lee theory of BCS superconductivity to elucidate the nonperturbative nature of the superconducting phase transition in terms of the distribution of zeros of the partition function. We show that the number of roots of the partition function is directly related to the superconducting gap induced by the Fermi-surface instability. In particular, we demonstrate that the Yang-Lee zeros of the partition function are distributed on a semicircle in the complex plane of the interaction strength, where the superconducting phase transition occurs at the edge of the distribution.
A previous study \cite{Sumaryada:2007uu} on Fisher zeros in pairing fields focuses on a finite-temperature phase transition by making the temperature complex. In contrast, we focus on the zero-temperature quantum phase transition by making the interaction strength complex.  

Furthermore, we employ a renormalization group (RG) to investigate the 
universality of the distribution of Yang-Lee zeros
for a generic quantum many-body system with a marginal coupling. In particular, we show the semicircle theorem: Yang-Lee zeros in a quantum many-body system with a marginally relevant coupling are distributed on a semicircle
in the complex interaction plane, in contrast to a full circle of the original Lee-Yang circle
theorem \cite{PhysRev.87.404,PhysRev.87.410}. This general theorem demonstrates that the geometric shape of the distribution of Yang-Lee zeros is directly connected to the existence of the Fermi-surface instability.

Last, we investigate the nonunitary criticality in BCS superconductivity which originates from the Yang-Lee singularity. By determining the critical exponents, we show that the nonunitary singularity belongs to a universality
class distinct from that of the Hermitian superconducting phase transition. The unconventional quantum critical phenomena are caused by exceptional points where a nonanalytic excitation spectrum emerges near the Fermi surface \cite{Yamamoto2019}.

\emph{Yang-Lee singularity in superconductivity}.---We consider a three-dimensional
BCS model \cite{Yamamoto2019}\footnote{Note that in our definition, $U_{R}>0\:(<0)$ represents attractive
(repulsive) interaction.} 
\begin{equation}
H=\sum_{\boldsymbol{k}\sigma}\xi_{\boldsymbol{k}}c_{\boldsymbol{k}\sigma}^{\dagger}c_{\boldsymbol{k}\sigma}-\frac{U}{N}\sum_{\bm{k},\bm{k}'}{}^{'}c_{\bm{k}\uparrow}^{\dagger}c_{\bm{-k}\downarrow}^{\dagger}c_{\bm{-k}'\downarrow}c_{\bm{k}'\uparrow},\label{eq:non-Hermitian}
\end{equation}
where $\xi_{\boldsymbol{k}}=\epsilon_{\bm{k}}-\mu$ is the single-particle
energy measured from the chemical potential $\mu$, $\sigma=\uparrow,\downarrow$
is the spin index, and $U=U_{R}+iU_{I}$ is the complex-valued
interaction strength. 
The creation and annihilation operators of an electron with momentum $\bm{k}$ and spin $\sigma$ are denoted as $c_{\bm{k}\sigma}^{\dagger}$
and $c_{\bm{k}\sigma}$, respectively. The prime in $\sum_{\bm{k}}^{'}$
indicates that the sum over $\bm{k}$ is restricted to $|\xi_{\boldsymbol{k}}|<\omega_{D}$,
where $\omega_{D}$ is the energy cutoff and $N$ is the number of momenta within this cutoff. Note that the non-Hermitian Hamiltonian (\ref{eq:non-Hermitian}) is used to investigate Yang-Lee zeros of closed systems as opposed to open systems in Ref. \cite{Yamamoto2019}. 

A non-Hermitian generalization of the BCS theory is made in Ref. \cite{Yamamoto2019}, where the mean-field BCS Hamiltonian
is given by $H_{\mathrm{MF}}=\sum_{\boldsymbol{k}\sigma}\xi_{\boldsymbol{k}}c_{\boldsymbol{k}\sigma}^{\dagger}c_{\boldsymbol{k}\sigma}+\sum_{\bm{k}}^{'}[\bar{\Delta}_0c_{-\bm{k}\downarrow}c_{\bm{k}\uparrow}+\Delta_0 c_{\bm{k}\uparrow}^{\dagger}c_{-\bm{k}\downarrow}^{\dagger}]+\frac{N}{U}\bar{\Delta}_0\Delta_0$,
with the superconducting gaps $\Delta_0=-\frac{U}{N}\sum_{\boldsymbol{k}L}^{'}\langle c_{-\boldsymbol{k}\downarrow}c_{\boldsymbol{k}\uparrow}\rangle_{R}$
and $\bar{\Delta}_0=-\frac{U}{N}\sum_{\boldsymbol{k}L}^{'}\langle c_{\boldsymbol{k}\uparrow}^{\dagger}c_{-\boldsymbol{k}\downarrow}^{\dagger}\rangle_{R}$.
Here $_{L}\langle A\rangle_{R}:={}_{L}\langle\text{BCS}|A|\text{BCS}\rangle_{R}$,
and $|\text{BCS}\rangle_{R}$ and $|\text{BCS}\rangle_{L}$ are the
right and left ground states of the Hamiltonian $H_{\mathrm{MF}}$ given by \cite{Yamamoto2019} 
\begin{align}
|\text{BCS}\rangle_{R} & =\prod_{\bm{k}}(u_{\bm{k}}+v_{\bm{k}}c_{\boldsymbol{k}\uparrow}^{\dagger}c_{-\boldsymbol{k}\downarrow}^{\dagger})|0\rangle,\\
|\text{BCS}\rangle_{L} & =\prod_{\bm{k}}(u_{\bm{k}}^{*}+\bar{v}_{\bm{k}}^{*}c_{\boldsymbol{k}\uparrow}^{\dagger}c_{-\boldsymbol{k}\downarrow}^{\dagger})|0\rangle,
\end{align}
where $|0\rangle$ is the vacuum state for electrons, and $u_{\bm{k}},v_{\bm{k}}$,
and $\bar{v}_{\bm{k}}$ are complex coefficients subject to the normalization
condition $u_{\bm{k}}^{2}+v_{\bm{k}}\bar{v}_{\bm{k}}=1$. These coefficients
can be determined in a standard manner and given in Supplemental Material
\cite{SupplementaryMaterial}. Since the right and left ground states
are not the same, $\Delta_0\neq\bar{\Delta}_0^{*}$ and $\bar{v}_{\bm{k}}\neq v_{\bm{k}}^{*}$
in general. Here we take a gauge such that $\bar{\Delta}_0=\Delta_0$. The Bogoliubov energy spectrum $E_{\bm{k}}$ is given by
\cite{Yamamoto2019} 
$E_{\bm{k}}=\sqrt{\xi_{\bm{k}}^{2}+\Delta_{\bm{k}}^{2}}$,
where $\Delta_{\bm{k}}=\Delta_0\theta(\omega_{D}-|\xi_{\bm{k}}|)$
with $\theta(x)$ being the Heaviside unit-step function. It is worthwhile
to note that $\Delta_0$ is complex in general, so is the energy
$E_{\bm{k}}$. In the following, we assume that the density of states
$\rho_{0}$ in the energy shell is a constant. The gap $\Delta_0$
is then given by 
\begin{equation}
\Delta_0=\frac{\omega_{D}}{\text{sinh}\left(\frac{1}{\rho_{0}U}\right)},\label{eq:gap}
\end{equation}
which exhibits an essential singularity at $U=0$.

\begin{figure}
\includegraphics[width=1\columnwidth]{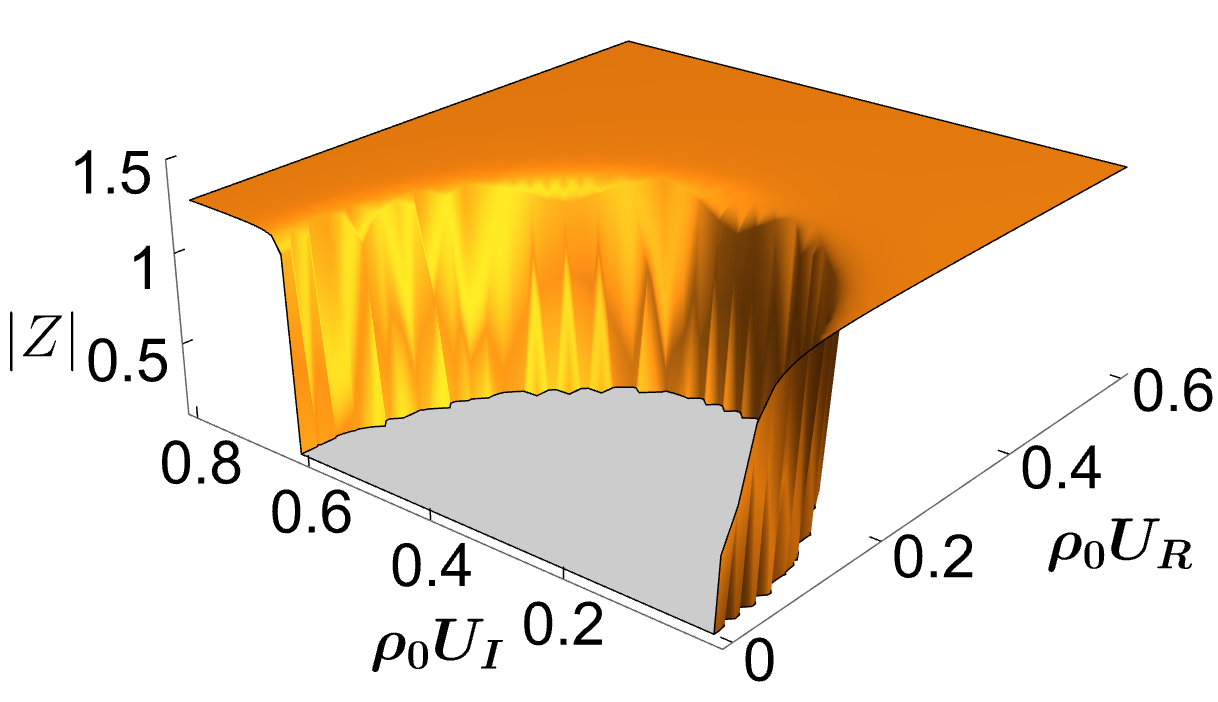}

\caption{Absolute value of the partition function $Z$ of the three-dimensional
BCS model as a function of the real and imaginary parts of the interaction
strength $U=U_{R}+iU_{I}$ in the zero-temperature limit. The boundary
along which the partition function vanishes is given by Eq. (\ref{eq:phase_transition}). In the gray region inside
the phase boundary, the value of the partition function is not shown
due to the breakdown of the mean-field approximation \cite{Yamamoto2019}.}

\label{Phase_transition_line}
\end{figure}

The partition function is given by
\begin{equation}
Z=\prod_{\bm{k},\sigma}Z_{\bm{k}}=\prod_{\bm{k},\sigma}(1+e^{-\beta E_{\bm{k}}}),\label{eq:analytical_expression_partition}
\end{equation}
whose absolute value is shown in Fig. \ref{Phase_transition_line}. Here $\beta$ is the inverse temperature. 
The Yang-Lee zeros of our system are defined by zeros of the
partition function in Eq. (\ref{eq:analytical_expression_partition})
where $\mathrm{Re}(E_{\bm{k}})=0$ and $\mathrm{Im}(\beta E_{\bm{k}})=(2n+1)\pi,n\in\mathbb{Z}$.
This condition is satisfied in the thermodynamic limit if $\mathrm{Re}\Delta_0=0$,
which agrees with the condition of phase transitions. It follows from this condition that the positions of Yang-Lee zeros satisify

\begin{equation}
(\rho_{0}\pi U_{R})^{2}+(\rho_{0}\pi U_{I}-1)^{2}=1,\;U_{R}>0.\label{eq:phase_transition}
\end{equation}
Note that these points coincide with the exceptional points where $H_{\mathrm{MF}}$
is not diagonalizable \cite{Yamamoto2019}. The Yang-Lee
zeros are distributed on a semicircle in the complex plane of the interaction strength $U$
depicted as the boundary of the gray region in Fig. \ref{Phase_transition_line}. In the yellow region in Fig. \ref{Phase_transition_line}, the energy spectrum $E_{\mathbf{k}}$ is gapped and $|Z|\to 1$ in the zero-temperature limit since $e^{-\beta E_{\bm{k}}}\to 0$ for all momenta $\bm{k}$. Note that the distribution of the Yang-Lee zeros touches the real axis at the origin, which is consistent with the fact that the superconducting phase transition occurs at the origin.

The essential singularity at the superconducting phase transition is directly linked to the number of roots $\chi$ of the partition function. According to Eq. (\ref{eq:analytical_expression_partition}), each factor in the partition function contributes to one root if and only if $\mathrm{Re}(E_{\bm{k}})=0$ and $\mathrm{Im}(\beta E_{\bm{k}})=(2n+1)\pi$ for some $n\in\mathbb{Z}$. Therefore, the number of roots of the partition function in the complex energy space is given by the number of integer $n$ satisfying $\mathrm{Im}(\beta E_{\bm{k}})=(2n+1)\pi$ for $ \mathrm{Im}(\beta E_{\bm{k}})\in [0, \beta \text{Im}(\Delta_0)]$ (see Eqs. (12)--(15) in Supplemental Material). From Eq. (\ref{eq:gap}), we have
\begin{equation}
\chi\simeq\frac{\beta\,\text{Im}(\Delta_0)}{\pi}=\frac{\beta\omega_{D}}{\pi\,\text{cosh}(\frac{U_{R}}{\rho_{0}|U|^{2}})}\label{eq:order}
\end{equation}
in the zero-temperature limit, where the spin-degeneracy factor of two is included. 
Near $U=0$, the gap takes the form of $\Delta_0=2\omega_{D}\exp(-\frac{1}{\rho_{0}U})$.
Since the phase boundary is tangent to the real axis where $U_{I}\ll U_{R}$,
we may put $U_{I}=0$ in Eq. (\ref{eq:order}) near the origin, obtaining
\begin{equation}
\chi\simeq\frac{\beta}{\pi}\Delta_0|_{U_I=0}.\label{eq:chivalue}
\end{equation}
Equation (\ref{eq:chivalue}) relates the number of roots $\chi$ of
the partition function 
to the superconducting gap $\Delta_0$ on the real axis. The 
condensation energy $\Delta E=F(\Delta_0)-F(0)$ \cite{Coleman:2015vz}, 
where $F$ is the free energy, can be obtained from the number of roots $\chi$ 
as \cite{SupplementaryMaterial}
\begin{equation}
	\Delta E\simeq-\frac{\pi^2N\rho_0}{2}\left(\frac{\chi}{\beta}\right)^2\propto\chi^2.
\end{equation}
For a general phase transition with spontaneous symmetry breaking and order parameter $\Delta_0$ with the dimension of energy, we have $\Delta E\propto\chi^2$ \footnote{In general we can write the Ginzberg-Landau free energy in terms of order parameter $\Delta_0$ as $F=a\Delta_0^2+b\Delta_0^4$. Near the origin we have $\Delta_0\to0$ and hence $F\propto\Delta_0^2\propto\chi^2$ holds.}. The condensation energy is thus directly related to the number of roots $\chi$. 

\emph{Semicircle theorem}.---Here, we show that the semicircular
distribution (\ref{eq:phase_transition}) of the Yang-Lee zeros is generic and universal in quantum many-body systems.
We consider a general canonical RG equation for a marginal
complex interaction: 
\begin{equation}
\frac{dV}{dt}=aV^{2}+bV^{3},\label{eq:RG_Flow_equation}
\end{equation}
where $V=V_{R}+iV_{I}\in\mathbb{C}$ is a dimensionless coupling strength
which can be taken as $V=\rho_{0}U$ in the present case, and $dt=-\frac{d\Xi}{\Xi}$
is the relative width of the high-energy shell which is to be integrated out in the Wilsonian RG with $\Xi$ being the energy cutoff. There are two finite fixed points in Eq. (\ref{eq:RG_Flow_equation}). One is $V=0$, which is trivial, and the other is $V=-\frac{a}{b}$, which is nontrivial. According to the stability of the nontrivial
fixed point, we can classify the RG-flow diagrams into two types depending on the sign of $b$.
The case with $b>0$ corresponds to an unstable nontrivial fixed point
in the Hermitian case and does not exhibit critical phenomena. The
other case with $b<0$ corresponds to a stable nontrivial fixed point
in the Hermitian case and is the only one that includes the critical line. The BCS model belongs to this case.
The RG-flow diagrams for these cases are shown in Supplemental
Material \cite{SupplementaryMaterial}. By applying the Wilsonian RG analysis
of the fermionic field theory \cite{Shankar1994}, the RG equation of
the BCS model up to the two-loop order including the self-energy correction
is written as \cite{SupplementaryMaterial} 
\begin{equation}
\frac{dV}{dt}=V^{2}-\frac{1}{2}V^{3}.\label{eq:rg_equation}
\end{equation}
From Eq. (\ref{eq:rg_equation}), we find $a=1$ and $b=-1/2$ in
the canonical RG equation (\ref{eq:RG_Flow_equation}). A similar RG
equation has been obtained for the non-Hermitian Kondo model \cite{Nakagawa2018}.
Note that the sign of the parameter $a$ does not influence the physics
of RG flows since we can reverse its sign by the replacement $V\to-V$.
For a system with $b<0$, there exists a critical line which separates
the trivial and nontrivial fixed points. Every point on the critical
line flows toward an infinite fixed point $(V_{R},V_{I})=(-a/(3b),\infty)$.
After integrating Eq. (\ref{eq:RG_Flow_equation}) and taking the
imaginary part, we obtain the critical line as
\begin{equation}
\frac{b\pi}{|a|}+\frac{V_{I}}{V_{R}^{2}+V_{I}^{2}}=\frac{b}{a}\arctan{\frac{V_{I}}{V_{R}}}+\frac{b}{a}\arctan{\left(-\frac{bV_{I}}{a+bV_{R}}\right)}.\label{critical_line_exact}
\end{equation}
Near the origin, Eq. (\ref{critical_line_exact}) can be expanded
as 
\begin{equation}
\frac{V_{I}}{V_{R}^{2}+V_{I}^{2}}+\frac{b\pi}{|a|}=0.\label{critical_line}
\end{equation}
The critical line specified by Eqs. (\ref{critical_line_exact}) and (\ref{critical_line})
is located in the right-half complex plane $V_{R}>0$ for $a>0$ and in the left-half complex plane $V_{R}<0$ for $a<0$. Note that the critical line (\ref{critical_line})
forms a semicircle for all $a\neq0$ and $b<0$. For the BCS model,
Eq. (\ref{critical_line}) reduces to $-U_{I}/[\rho_{0}(U_{R}^{2}+U_{I}^{2})]+\pi/2=0$,
which agrees with the mean-field phase boundary in Eq. (\ref{eq:phase_transition})
along which the Yang-Lee zeros are distributed. This RG result confirms the validity
of the mean-field results. 

The above analysis of general marginally interacting systems with $a\neq0$
and $b<0$ implies that the criticality associated with the Yang-Lee
zeros, if exists, can only take place on the semicircle (\ref{critical_line})
within the perturbative RG framework. This semicircle distribution
of Yang-Lee zeros is to be compared with the Lee-Yang circle theorem \cite{PhysRev.87.410}
where the zeros are distributed on a unit circle. The semicircle structure
arises from the marginal nature of the coupling strength that induces different
RG-flow behaviors between the left-half plane $V_{R}<0$ and the right-half plane $V_{R}>0$. In fact, such a feature lies at the heart of the Fermi-surface instability in superconductivity. 

This semicircle theorem indicates that the nonperturbative properties in the Fermi-surface instability is generally linked to the geometric shape of the distribution of Yang-Lee zeros. This semicircular distribution of Yang-Lee zeros should appear in diverse systems subject to Fermi-surface instabilities, such as the charge-density wave (CDW) and anisotropic Cooper pairing,
since they are described by similar RG equations with
marginal couplings \cite{Shankar1994}. In fact, systems with the
CDW instability can be described by a mean-field analysis similar
to the BCS theory \cite{Gersch2005,PhysRevB.103.045142,PhysRevB.72.195106}. 

\emph{Nonunitary critical phenomena in superconductivity.} ---The
Yang-Lee zeros in the complex plane are accompanied by
nonunitary critical phenomena in BCS superconductivity
with a complex-valued interaction.  
Remarkably, the criticality in the BCS model arises at every point on the phase boundary rather than at the edges alone as in the Ising model.

We now examine the critical exponents and the universality class of the Yang-Lee singularity. The correlation function

\begin{align}
C(\bm{x}) & =_{L}\langle c_{\sigma}^{\dagger}(\bm{x})c_{\sigma}(0)\rangle_{R}\nonumber \\
 & :=_{L}\langle\text{BCS}|c_{\sigma}^{\dagger}(\bm{x})c_{\sigma}(0)|\text{BCS}\rangle_{R}
\end{align}
can be calculated from the Fourier transformation
\begin{equation}
C(\bm{x})\simeq-\frac{1}{N}\sum_{\bm{k}}^ {}{'}\frac{\xi_{\bm{k}}}{2E_{\bm{k}}}e^{i\bm{k}\cdot\bm{x}}.\label{eq:correlation_Fourier}
\end{equation}
Here we restrict the sum over $\bm{k}$ to the energy shell since
we are concerned with the long-range behavior of the correlation function.
We expand $\xi_{\bm{k}}$ near the Fermi surface as $\xi_{{\bm{k}}}=v_{F}(k-k_{F})$,
where $v_{F}$ is the Fermi velocity, $k_{F}$ is the Fermi momentum
and $k=|\bm{k}|$. On the phase boundary (\ref{eq:phase_transition}),
the correlation function (\ref{eq:correlation_Fourier}) shows a
power-law decay as 
\begin{equation}
\lim_{x\rightarrow\infty}C(\bm{x})\simeq\frac{A(l)}{l^{3/2}}+i\frac{B(l)}{l^{3/2}}\propto x^{-3/2},\label{correlation}
\end{equation}
where $x=|\bm{x}|$, $l:=\frac{\text{Im}\Delta_0}{v_{F}}x$ is a
dimensionless length scale, and $A(l)$ and $B(l)$ are real functions
that oscillate with $l$ without decay (see Supplemental Material
\cite{SupplementaryMaterial} for details). The anomalous power of
$\frac{3}{2}$ arises from the exceptional points of the system, and should be compared with the power of 2 for the normal-metal
phase \cite{Sachdev:2011uj}. When the gap closes at the exceptional points, the
dispersion relation near the Fermi surface is given by

\begin{equation}
E_{\bm{k}}\simeq\sqrt{v_{F}^{2}k^{2}-(\text{Im}\Delta_0)^{2}}.
\end{equation}
Near the exceptional points $k_{E}:=\frac{\text{Im}\Delta_0}{v_{F}}$,
the dispersion relation reduces to $E_{\bm{k}}\sim\sqrt{k-k_{E}}$,
in sharp contrast with the Hermitian counterpart which exhibits 
a linear dispersion relation near a gapless point. It is this square-root
excitation spectrum that induces the anomalous decay of the correlation
function near the phase boundary. From the correlation function (\ref{correlation}),
we find the anomalous dimension $\eta=1/2$ from $C(\bm{x})\propto x^{-D+2-\eta}$
on the phase boundary, where $D$ is the dimension of the system \cite{Sachdev:2011uj}.

The correlation function decays exponentially near the phase boundary.
If we shift $U$ by an infinitesimal amount $\delta U$ along the
real axis from the phase boundary, the correlation function can also
be calculated from Eq. (\ref{eq:correlation_Fourier}), giving

\begin{equation}
\lim_{x\rightarrow\infty}C(\bm{x})\propto[A(l)+iB(l)]\frac{\text{exp}\left(-\frac{l}{\xi}\right)}{l^{3/2}},\label{eq:exponential_decay}
\end{equation}
where the correlation length $\xi\propto(\rho_{0}\delta U)^{-1}$
diverges on the phase boundary, and hence we obtain the critical exponent
$\nu=1$ from $\xi\propto(\delta U)^{-\nu}$ \cite{Sachdev:2011uj}
(see Supplemental Material \cite{SupplementaryMaterial} for the
derivation). Near the phase boundary, the dynamical critical exponent
$z$ is defined as 
\begin{equation}
\text{Re}\Delta_0\propto\xi^{-z}.
\end{equation}
From the expression of $\Delta_0$ in Eq. (\ref{eq:gap}), we find
that $\text{Re}\Delta_0\propto\xi^{-1}\propto\delta U$. Hence 
we have $z=1$.

The correlation length in the Hermitian case takes the
form of

\begin{equation}
\xi\propto\text{exp}\left(\frac{1}{\rho_{0}\delta U}\right).\label{eq:xi}
\end{equation}
This behavior is distinct from that of the quantum phase transition
in the non-Hermitian case since $\xi^{-1}$ in Eq. (\ref{eq:xi})
cannot be expanded as a power series of $\rho_{0}\delta U$, indicating that the exceptional points lead to a distinct 
universality class in the non-Hermitian system.

We next consider the pair correlation function 
\begin{align}
 & \rho_{2}(\bm{r}_{1}\sigma_{1},\bm{r}_{2}\sigma_{2};\bm{r'_{1}}\sigma'_{1},\bm{r'}_{2}\sigma'_{2})\nonumber \\
 & =_{L}\langle c_{\sigma_{1}}^{\dagger}(\bm{r}_{1})c_{\sigma_{2}}^{\dagger}(\bm{r}_{2})c_{\sigma'_{2}}(\bm{r'}_{2})c_{\sigma'_{1}}(\bm{r'}_{1})\rangle_{R},
\end{align}
where $(\bm{r}_{1}\sigma_{1},\bm{r}_{2}\sigma_{2})$ and $(\bm{r'_{1}}\sigma'_{1},\bm{r'}_{2}\sigma'_{2})$
are the positions and spins of electrons that form Cooper pairs.
Setting $\bm{r}_{1}=\bm{r}_{2}=\bm{R}$ and $\bm{r'_{1}}=\bm{r'_2}=0$
and taking the limit $|\bm{R}|\rightarrow\infty$, we find that $\rho_{2}$ converges to a nonzero value
on the phase boundary as

\begin{equation}
\lim_{R\rightarrow\infty}\rho_{2}(\bm{R}\uparrow,\bm{R}\downarrow;0\downarrow,0\uparrow)=-\frac{(\text{Im}\Delta_0)^{2}}{U^{2}}\neq0.\label{eq:pairing}
\end{equation}
This nonvanishing pair correlation function is characteristic of
nonunitary critical phenomena, where the correlation function of the
order parameter may diverge at long distance \cite{Fisher:1978vn}.
We can also use Eq. (\ref{eq:pairing}) to define the critical
exponent $\delta$ as 
\begin{equation}
\lim_{R\rightarrow\infty}\rho_{2}(\bm{R}\uparrow,\bm{R}\downarrow;0\downarrow,0\uparrow)\propto|\bm{R}|^{-\delta}.
\end{equation}
We have $\delta=0$ here, which is also unique to the nonunitary critical
phenomena.

The compressibility also shows critical behavior associated with the Yang-Lee singularity.
By analyzing the compressibility $\kappa=\frac{\partial^{2}F}{\partial\mu^{2}}$
near the phase boundary where $F=-(1/\beta)\log Z$ is the free energy
of Bogoliubov quasiparticles, we have 
\begin{equation}
\kappa=-N\int_{-\omega_{D}}^{\omega_{D}}\rho_{0}d\xi_{\bm{k}}\frac{\Delta_0^{2}}{(\xi_{\bm{k}}^{2}+\Delta_0^{2})^{3/2}}.
\end{equation}
On the phase boundary (\ref{eq:phase_transition}), the compressibility
$\kappa$ diverges. Therefore, we define another critical exponent
$\zeta$ near the phase boundary as 
\begin{equation}
\kappa\propto(\delta U)^{-\zeta}\,,
\end{equation}
with $\zeta=1/2$ in this system. This critical behavior also arises
from the nonanalytic square-root dispersion relation near the exceptional
points. In fact, the critical exponents $\eta$ and $\zeta$ are equal to each other 
for a general fractional-power dispersion relation $(k-k_{E})^{1/n}$,
which includes the case of higher-order exceptional points \cite{SupplementaryMaterial}. 

These power-law behaviors in the nonunitary
critical phenomena constitute a new Yang-Lee universality class distinct from that of the Yang-Lee edge singularity \cite{Fisher:1978vn}.
From the RG analysis, each point on the phase
boundary (\ref{critical_line_exact}) except for the origin flows
to $(\rho_0U_R,\rho_0U_I)=(\frac{2}{3},\infty)$, while the origin remains invariant in
the RG flow. Hence, the points on the phase boundary except for the
origin represent a universality class different from that at the
origin.

\emph{Conclusion}.---We have investigated the Yang-Lee
zeros in BCS superconductivity and found that the Yang-Lee zeros are distributed
on the semicircular phase boundary in the complex plane of the interaction
strength. We find that the nonperturbative nature of the order parameter and thermodynamic quantities are directly connected to the number of roots of the partition function, which allows us to understand superconducting quantum phase transitions from the analytic property of the partition function. We have performed the RG analysis
of generic many-body fermionic systems with marginal interactions and shown that the semicircle distribution of Yang-Lee zeros is a universal phenomenon in Fermi systems. 
We have also explored the Yang-Lee critical behavior and obtained critical exponents of the nonunitary criticality.

The Yang-Lee zeros and the corresponding singularity studied in this
Letter are not only an interesting mathematical property but can also be tested experimentally. In fact, the non-Hermitian BCS model can be realized in
open quantum systems \cite{Yamamoto2019,PhysRevA.103.013724}. The
complex-valued interaction strength describes the effect of two-body
loss in ultracold atoms. For example, inelastic two-body losses can
be induced by utilizing Feshbach resonances \cite{PhysRevLett.115.265301,PhysRevLett.115.265302,Zhang2015}
or photoassociation \cite{doi:10.1126/sciadv.1701513,Honda2023}.
Since the dissipation in these cases only involves atomic loss, the eigenvalue spectrum and the exceptional points of the Lindblad equation including the jump terms are the same as those of the corresponding non-Hermitian Hamiltonian regardless of the jump terms \cite{PhysRevLett.126.110404}. Hence, we believe the nonunitary critical phenomena introduced in this Letter should also be observed in open quantum systems. 

While we have focused on the quantum phase transition, it is worthwhile
to investigate how the Yang-Lee singularity is connected to a superconducting
phase transition at finite temperature. We also expect that Yang-Lee
zeros can emerge in other non-Hermitian many-body systems such as
a non-Hermitian Bose-Hubbard model \cite{PhysRevA.94.053615}.

We are grateful to Yuto Ashida, Kazuaki Takasan, Norifumi Matsumoto,
Kohei Kawabata, Xin Chen and Xuanzhao Gao for fruitful discussion.
H. L. is supported by Forefront Physics and Mathematics Program to
Drive Transformation (FoPM), a World-leading Innovative Graduate Study
(WINGS) Program, the University of Tokyo. X. Y. is supported by the
Munich Quantum Valley, which is supported by the Bavarian state government
with funds from the Hightech Agenda Bayern Plus. M.N. is supported
by JSPS KAKENHI Grant No. JP20K14383. M.U. is supported by JSPS KAKENHI
Grant No. JP22H01152 and the CREST program ``Quantum Frontiers" of JST (Grand No. JPMJCR23I1).
\bibliographystyle{apsrev4-2}
\bibliography{MyCollection}

\begin{thebibliography}{50}%
\makeatletter
\providecommand \@ifxundefined [1]{%
 \@ifx{#1\undefined}
}%
\providecommand \@ifnum [1]{%
 \ifnum #1\expandafter \@firstoftwo
 \else \expandafter \@secondoftwo
 \fi
}%
\providecommand \@ifx [1]{%
 \ifx #1\expandafter \@firstoftwo
 \else \expandafter \@secondoftwo
 \fi
}%
\providecommand \natexlab [1]{#1}%
\providecommand \enquote  [1]{``#1''}%
\providecommand \bibnamefont  [1]{#1}%
\providecommand \bibfnamefont [1]{#1}%
\providecommand \citenamefont [1]{#1}%
\providecommand \href@noop [0]{\@secondoftwo}%
\providecommand \href [0]{\begingroup \@sanitize@url \@href}%
\providecommand \@href[1]{\@@startlink{#1}\@@href}%
\providecommand \@@href[1]{\endgroup#1\@@endlink}%
\providecommand \@sanitize@url [0]{\catcode `\\12\catcode `\$12\catcode
  `\&12\catcode `\#12\catcode `\^12\catcode `\_12\catcode `\%12\relax}%
\providecommand \@@startlink[1]{}%
\providecommand \@@endlink[0]{}%
\providecommand \url  [0]{\begingroup\@sanitize@url \@url }%
\providecommand \@url [1]{\endgroup\@href {#1}{\urlprefix }}%
\providecommand \urlprefix  [0]{URL }%
\providecommand \Eprint [0]{\href }%
\providecommand \doibase [0]{https://doi.org/}%
\providecommand \selectlanguage [0]{\@gobble}%
\providecommand \bibinfo  [0]{\@secondoftwo}%
\providecommand \bibfield  [0]{\@secondoftwo}%
\providecommand \translation [1]{[#1]}%
\providecommand \BibitemOpen [0]{}%
\providecommand \bibitemStop [0]{}%
\providecommand \bibitemNoStop [0]{.\EOS\space}%
\providecommand \EOS [0]{\spacefactor3000\relax}%
\providecommand \BibitemShut  [1]{\csname bibitem#1\endcsname}%
\let\auto@bib@innerbib\@empty
\bibitem [{\citenamefont {Yang}\ and\ \citenamefont
  {Lee}(1952)}]{PhysRev.87.404}%
  \BibitemOpen
  \bibfield  {author} {\bibinfo {author} {\bibfnamefont {C.~N.}\ \bibnamefont
  {Yang}}\ and\ \bibinfo {author} {\bibfnamefont {T.~D.}\ \bibnamefont {Lee}},\
  }\href {https://doi.org/10.1103/PhysRev.87.404} {\bibfield  {journal}
  {\bibinfo  {journal} {Phys. Rev.}\ }\textbf {\bibinfo {volume} {87}},\
  \bibinfo {pages} {404} (\bibinfo {year} {1952})}\BibitemShut {NoStop}%
\bibitem [{\citenamefont {Lee}\ and\ \citenamefont
  {Yang}(1952)}]{PhysRev.87.410}%
  \BibitemOpen
  \bibfield  {author} {\bibinfo {author} {\bibfnamefont {T.~D.}\ \bibnamefont
  {Lee}}\ and\ \bibinfo {author} {\bibfnamefont {C.~N.}\ \bibnamefont {Yang}},\
  }\href {https://doi.org/10.1103/PhysRev.87.410} {\bibfield  {journal}
  {\bibinfo  {journal} {Phys. Rev.}\ }\textbf {\bibinfo {volume} {87}},\
  \bibinfo {pages} {410} (\bibinfo {year} {1952})}\BibitemShut {NoStop}%
\bibitem [{\citenamefont {Fisher}(1978)}]{Fisher:1978vn}%
  \BibitemOpen
  \bibfield  {author} {\bibinfo {author} {\bibfnamefont {M.~E.}\ \bibnamefont
  {Fisher}},\ }\href {https://doi.org/10.1103/PhysRevLett.40.1610} {\bibfield
  {journal} {\bibinfo  {journal} {Phys. Rev. Lett.}\ }\textbf {\bibinfo
  {volume} {40}},\ \bibinfo {pages} {1610} (\bibinfo {year}
  {1978})}\BibitemShut {NoStop}%
\bibitem [{\citenamefont {Kurtze}\ and\ \citenamefont
  {Fisher}(1979)}]{Kurtze:1979wb}%
  \BibitemOpen
  \bibfield  {author} {\bibinfo {author} {\bibfnamefont {D.~A.}\ \bibnamefont
  {Kurtze}}\ and\ \bibinfo {author} {\bibfnamefont {M.~E.}\ \bibnamefont
  {Fisher}},\ }\href {https://doi.org/10.1103/PhysRevB.20.2785} {\bibfield
  {journal} {\bibinfo  {journal} {Phys. Rev. B}\ }\textbf {\bibinfo {volume}
  {20}},\ \bibinfo {pages} {2785} (\bibinfo {year} {1979})}\BibitemShut
  {NoStop}%
\bibitem [{\citenamefont {Fisher}(1980)}]{10.1143/PTP.69.14}%
  \BibitemOpen
  \bibfield  {author} {\bibinfo {author} {\bibfnamefont {M.~E.}\ \bibnamefont
  {Fisher}},\ }\href {https://doi.org/10.1143/PTP.69.14} {\bibfield  {journal}
  {\bibinfo  {journal} {Prog. Theor. Phys. Suppl.}\ }\textbf {\bibinfo {volume}
  {69}},\ \bibinfo {pages} {14} (\bibinfo {year} {1980})}\BibitemShut {NoStop}%
\bibitem [{\citenamefont {Cardy}(1985)}]{Cardy:1985ub}%
  \BibitemOpen
  \bibfield  {author} {\bibinfo {author} {\bibfnamefont {J.~L.}\ \bibnamefont
  {Cardy}},\ }\href {https://doi.org/10.1103/PhysRevLett.54.1354} {\bibfield
  {journal} {\bibinfo  {journal} {Phys. Rev. Lett.}\ }\textbf {\bibinfo
  {volume} {54}},\ \bibinfo {pages} {1354} (\bibinfo {year}
  {1985})}\BibitemShut {NoStop}%
\bibitem [{\citenamefont {Cardy}\ and\ \citenamefont
  {Mussardo}(1989)}]{Cardy:1989uo}%
  \BibitemOpen
  \bibfield  {author} {\bibinfo {author} {\bibfnamefont {J.~L.}\ \bibnamefont
  {Cardy}}\ and\ \bibinfo {author} {\bibfnamefont {G.}~\bibnamefont
  {Mussardo}},\ }\href
  {https://doi.org/https://doi.org/10.1016/0370-2693(89)90818-6} {\bibfield
  {journal} {\bibinfo  {journal} {Phys. Lett. B}\ }\textbf {\bibinfo {volume}
  {225}},\ \bibinfo {pages} {275} (\bibinfo {year} {1989})}\BibitemShut
  {NoStop}%
\bibitem [{\citenamefont {Zamolodchikov}(1991)}]{Zamolodchikov:1991tl}%
  \BibitemOpen
  \bibfield  {author} {\bibinfo {author} {\bibfnamefont {A.~B.}\ \bibnamefont
  {Zamolodchikov}},\ }\href
  {https://doi.org/https://doi.org/10.1016/0550-3213(91)90207-E} {\bibfield
  {journal} {\bibinfo  {journal} {Nucl. Phys.}\ }\textbf {\bibinfo {volume}
  {B348}},\ \bibinfo {pages} {619} (\bibinfo {year} {1991})}\BibitemShut
  {NoStop}%
\bibitem [{\citenamefont {Bena}\ \emph {et~al.}(2005)\citenamefont {Bena},
  \citenamefont {Droz},\ and\ \citenamefont {Lipowski}}]{BENA2005}%
  \BibitemOpen
  \bibfield  {author} {\bibinfo {author} {\bibfnamefont {I.}~\bibnamefont
  {Bena}}, \bibinfo {author} {\bibfnamefont {M.}~\bibnamefont {Droz}},\ and\
  \bibinfo {author} {\bibfnamefont {A.}~\bibnamefont {Lipowski}},\ }\href
  {https://doi.org/10.1142/S0217979205032759} {\bibfield  {journal} {\bibinfo
  {journal} {Int. J. Mod. Phys. B}\ }\textbf {\bibinfo {volume} {19}},\
  \bibinfo {pages} {4269} (\bibinfo {year} {2005})}\BibitemShut {NoStop}%
\bibitem [{\citenamefont {Fisher}(1965)}]{Fisher1965}%
  \BibitemOpen
  \bibfield  {author} {\bibinfo {author} {\bibfnamefont {M.~E.}\ \bibnamefont
  {Fisher}},\ }\href {https://books.google.de/books?id=cBisnQEACAAJ} {\emph
  {\bibinfo {title} {The Nature of Critical Points}}}\ (\bibinfo  {publisher}
  {University of Colorado Press},\ \bibinfo {year} {1965})\BibitemShut
  {NoStop}%
\bibitem [{\citenamefont {Simon}\ and\ \citenamefont
  {Griffiths}(1973)}]{Simon:1973tr}%
  \BibitemOpen
  \bibfield  {author} {\bibinfo {author} {\bibfnamefont {B.}~\bibnamefont
  {Simon}}\ and\ \bibinfo {author} {\bibfnamefont {R.~B.}\ \bibnamefont
  {Griffiths}},\ }\href {https://doi.org/10.1007/BF01645626} {\bibfield
  {journal} {\bibinfo  {journal} {Commun. Math. Phys.}\ }\textbf {\bibinfo
  {volume} {33}},\ \bibinfo {pages} {145} (\bibinfo {year} {1973})}\BibitemShut
  {NoStop}%
\bibitem [{\citenamefont {Newman}(1974)}]{Newman:1974wi}%
  \BibitemOpen
  \bibfield  {author} {\bibinfo {author} {\bibfnamefont {C.~M.}\ \bibnamefont
  {Newman}},\ }\href {https://doi.org/https://doi.org/10.1002/cpa.3160270203}
  {\bibfield  {journal} {\bibinfo  {journal} {Comm. Pure Appl. Math.}\ }\textbf
  {\bibinfo {volume} {27}},\ \bibinfo {pages} {143} (\bibinfo {year}
  {1974})}\BibitemShut {NoStop}%
\bibitem [{\citenamefont {Lieb}\ and\ \citenamefont
  {Sokal}(1981)}]{Lieb:1981vb}%
  \BibitemOpen
  \bibfield  {author} {\bibinfo {author} {\bibfnamefont {E.~H.}\ \bibnamefont
  {Lieb}}\ and\ \bibinfo {author} {\bibfnamefont {A.~D.}\ \bibnamefont
  {Sokal}},\ }\href {https://doi.org/10.1007/BF01213009} {\bibfield  {journal}
  {\bibinfo  {journal} {Commun. Math. Phys.}\ }\textbf {\bibinfo {volume}
  {80}},\ \bibinfo {pages} {153} (\bibinfo {year} {1981})}\BibitemShut
  {NoStop}%
\bibitem [{\citenamefont {Kortman}\ and\ \citenamefont
  {Griffiths}(1971)}]{Kortman:1971tw}%
  \BibitemOpen
  \bibfield  {author} {\bibinfo {author} {\bibfnamefont {P.~J.}\ \bibnamefont
  {Kortman}}\ and\ \bibinfo {author} {\bibfnamefont {R.~B.}\ \bibnamefont
  {Griffiths}},\ }\href {https://doi.org/10.1103/PhysRevLett.27.1439}
  {\bibfield  {journal} {\bibinfo  {journal} {Phys. Rev. Lett.}\ }\textbf
  {\bibinfo {volume} {27}},\ \bibinfo {pages} {1439} (\bibinfo {year}
  {1971})}\BibitemShut {NoStop}%
\bibitem [{\citenamefont {Biskup}\ \emph {et~al.}(2000)\citenamefont {Biskup},
  \citenamefont {Borgs}, \citenamefont {Chayes}, \citenamefont {Kleinwaks},\
  and\ \citenamefont {Koteck\'y}}]{PhysRevLett.84.4794}%
  \BibitemOpen
  \bibfield  {author} {\bibinfo {author} {\bibfnamefont {M.}~\bibnamefont
  {Biskup}}, \bibinfo {author} {\bibfnamefont {C.}~\bibnamefont {Borgs}},
  \bibinfo {author} {\bibfnamefont {J.~T.}\ \bibnamefont {Chayes}}, \bibinfo
  {author} {\bibfnamefont {L.~J.}\ \bibnamefont {Kleinwaks}},\ and\ \bibinfo
  {author} {\bibfnamefont {R.}~\bibnamefont {Koteck\'y}},\ }\href
  {https://doi.org/10.1103/PhysRevLett.84.4794} {\bibfield  {journal} {\bibinfo
   {journal} {Phys. Rev. Lett.}\ }\textbf {\bibinfo {volume} {84}},\ \bibinfo
  {pages} {4794} (\bibinfo {year} {2000})}\BibitemShut {NoStop}%
\bibitem [{\citenamefont {Arndt}(2000)}]{PhysRevLett.84.814}%
  \BibitemOpen
  \bibfield  {author} {\bibinfo {author} {\bibfnamefont {P.~F.}\ \bibnamefont
  {Arndt}},\ }\href {https://doi.org/10.1103/PhysRevLett.84.814} {\bibfield
  {journal} {\bibinfo  {journal} {Phys. Rev. Lett.}\ }\textbf {\bibinfo
  {volume} {84}},\ \bibinfo {pages} {814} (\bibinfo {year} {2000})}\BibitemShut
  {NoStop}%
\bibitem [{\citenamefont {Blythe}\ and\ \citenamefont
  {Evans}(2002)}]{PhysRevLett.89.080601}%
  \BibitemOpen
  \bibfield  {author} {\bibinfo {author} {\bibfnamefont {R.~A.}\ \bibnamefont
  {Blythe}}\ and\ \bibinfo {author} {\bibfnamefont {M.~R.}\ \bibnamefont
  {Evans}},\ }\href {https://doi.org/10.1103/PhysRevLett.89.080601} {\bibfield
  {journal} {\bibinfo  {journal} {Phys. Rev. Lett.}\ }\textbf {\bibinfo
  {volume} {89}},\ \bibinfo {pages} {080601} (\bibinfo {year}
  {2002})}\BibitemShut {NoStop}%
\bibitem [{\citenamefont {Lee}(2013)}]{PhysRevLett.110.248101}%
  \BibitemOpen
  \bibfield  {author} {\bibinfo {author} {\bibfnamefont {J.}~\bibnamefont
  {Lee}},\ }\href {https://doi.org/10.1103/PhysRevLett.110.248101} {\bibfield
  {journal} {\bibinfo  {journal} {Phys. Rev. Lett.}\ }\textbf {\bibinfo
  {volume} {110}},\ \bibinfo {pages} {248101} (\bibinfo {year}
  {2013})}\BibitemShut {NoStop}%
\bibitem [{\citenamefont {von Gehlen}(1991)}]{Gehlen_1991}%
  \BibitemOpen
  \bibfield  {author} {\bibinfo {author} {\bibfnamefont {G.}~\bibnamefont {von
  Gehlen}},\ }\href {https://doi.org/10.1088/0305-4470/24/22/021} {\bibfield
  {journal} {\bibinfo  {journal} {J. Phys. A}\ }\textbf {\bibinfo {volume}
  {24}},\ \bibinfo {pages} {5371} (\bibinfo {year} {1991})}\BibitemShut
  {NoStop}%
\bibitem [{\citenamefont {Sumaryada}\ and\ \citenamefont
  {Volya}(2007)}]{Sumaryada:2007uu}%
  \BibitemOpen
  \bibfield  {author} {\bibinfo {author} {\bibfnamefont {T.}~\bibnamefont
  {Sumaryada}}\ and\ \bibinfo {author} {\bibfnamefont {A.}~\bibnamefont
  {Volya}},\ }\href {https://doi.org/10.1103/PhysRevC.76.024319} {\bibfield
  {journal} {\bibinfo  {journal} {Phys. Rev. C}\ }\textbf {\bibinfo {volume}
  {76}},\ \bibinfo {pages} {024319} (\bibinfo {year} {2007})}\BibitemShut
  {NoStop}%
\bibitem [{\citenamefont {Abraham}\ \emph {et~al.}(1996)\citenamefont
  {Abraham}, \citenamefont {Barbour}, \citenamefont {Cullen}, \citenamefont
  {Klepfish}, \citenamefont {Pike},\ and\ \citenamefont
  {Sarkar}}]{PhysRevB.53.7704}%
  \BibitemOpen
  \bibfield  {author} {\bibinfo {author} {\bibfnamefont {E.}~\bibnamefont
  {Abraham}}, \bibinfo {author} {\bibfnamefont {I.~M.}\ \bibnamefont
  {Barbour}}, \bibinfo {author} {\bibfnamefont {P.~H.}\ \bibnamefont {Cullen}},
  \bibinfo {author} {\bibfnamefont {E.~G.}\ \bibnamefont {Klepfish}}, \bibinfo
  {author} {\bibfnamefont {E.~R.}\ \bibnamefont {Pike}},\ and\ \bibinfo
  {author} {\bibfnamefont {S.}~\bibnamefont {Sarkar}},\ }\href
  {https://doi.org/10.1103/PhysRevB.53.7704} {\bibfield  {journal} {\bibinfo
  {journal} {Phys. Rev. B}\ }\textbf {\bibinfo {volume} {53}},\ \bibinfo
  {pages} {7704} (\bibinfo {year} {1996})}\BibitemShut {NoStop}%
\bibitem [{\citenamefont {Matsumoto}\ \emph {et~al.}(2022)\citenamefont
  {Matsumoto}, \citenamefont {Nakagawa},\ and\ \citenamefont
  {Ueda}}]{Matsumoto2020}%
  \BibitemOpen
  \bibfield  {author} {\bibinfo {author} {\bibfnamefont {N.}~\bibnamefont
  {Matsumoto}}, \bibinfo {author} {\bibfnamefont {M.}~\bibnamefont
  {Nakagawa}},\ and\ \bibinfo {author} {\bibfnamefont {M.}~\bibnamefont
  {Ueda}},\ }\href {https://doi.org/10.1103/PhysRevResearch.4.033250}
  {\bibfield  {journal} {\bibinfo  {journal} {Phys. Rev. Res.}\ }\textbf
  {\bibinfo {volume} {4}},\ \bibinfo {pages} {033250} (\bibinfo {year}
  {2022})}\BibitemShut {NoStop}%
\bibitem [{\citenamefont {Kist}\ \emph {et~al.}(2021)\citenamefont {Kist},
  \citenamefont {Lado},\ and\ \citenamefont
  {Flindt}}]{PhysRevResearch.3.033206}%
  \BibitemOpen
  \bibfield  {author} {\bibinfo {author} {\bibfnamefont {T.}~\bibnamefont
  {Kist}}, \bibinfo {author} {\bibfnamefont {J.~L.}\ \bibnamefont {Lado}},\
  and\ \bibinfo {author} {\bibfnamefont {C.}~\bibnamefont {Flindt}},\ }\href
  {https://doi.org/10.1103/PhysRevResearch.3.033206} {\bibfield  {journal}
  {\bibinfo  {journal} {Phys. Rev. Res.}\ }\textbf {\bibinfo {volume} {3}},\
  \bibinfo {pages} {033206} (\bibinfo {year} {2021})}\BibitemShut {NoStop}%
\bibitem [{\citenamefont {Vecsei}\ \emph {et~al.}(2022)\citenamefont {Vecsei},
  \citenamefont {Lado},\ and\ \citenamefont {Flindt}}]{PhysRevB.106.054402}%
  \BibitemOpen
  \bibfield  {author} {\bibinfo {author} {\bibfnamefont {P.~M.}\ \bibnamefont
  {Vecsei}}, \bibinfo {author} {\bibfnamefont {J.~L.}\ \bibnamefont {Lado}},\
  and\ \bibinfo {author} {\bibfnamefont {C.}~\bibnamefont {Flindt}},\ }\href
  {https://doi.org/10.1103/PhysRevB.106.054402} {\bibfield  {journal} {\bibinfo
   {journal} {Phys. Rev. B}\ }\textbf {\bibinfo {volume} {106}},\ \bibinfo
  {pages} {054402} (\bibinfo {year} {2022})}\BibitemShut {NoStop}%
\bibitem [{\citenamefont {Gnatenko}\ \emph {et~al.}(2017)\citenamefont
  {Gnatenko}, \citenamefont {Kargol},\ and\ \citenamefont
  {Tkachuk}}]{PhysRevE.96.032116}%
  \BibitemOpen
  \bibfield  {author} {\bibinfo {author} {\bibfnamefont {K.~P.}\ \bibnamefont
  {Gnatenko}}, \bibinfo {author} {\bibfnamefont {A.}~\bibnamefont {Kargol}},\
  and\ \bibinfo {author} {\bibfnamefont {V.~M.}\ \bibnamefont {Tkachuk}},\
  }\href {https://doi.org/10.1103/PhysRevE.96.032116} {\bibfield  {journal}
  {\bibinfo  {journal} {Phys. Rev. E}\ }\textbf {\bibinfo {volume} {96}},\
  \bibinfo {pages} {032116} (\bibinfo {year} {2017})}\BibitemShut {NoStop}%
\bibitem [{\citenamefont {Peotta}\ \emph {et~al.}(2021)\citenamefont {Peotta},
  \citenamefont {Brange}, \citenamefont {Deger}, \citenamefont {Ojanen},\ and\
  \citenamefont {Flindt}}]{PhysRevX.11.041018}%
  \BibitemOpen
  \bibfield  {author} {\bibinfo {author} {\bibfnamefont {S.}~\bibnamefont
  {Peotta}}, \bibinfo {author} {\bibfnamefont {F.}~\bibnamefont {Brange}},
  \bibinfo {author} {\bibfnamefont {A.}~\bibnamefont {Deger}}, \bibinfo
  {author} {\bibfnamefont {T.}~\bibnamefont {Ojanen}},\ and\ \bibinfo {author}
  {\bibfnamefont {C.}~\bibnamefont {Flindt}},\ }\href
  {https://doi.org/10.1103/PhysRevX.11.041018} {\bibfield  {journal} {\bibinfo
  {journal} {Phys. Rev. X}\ }\textbf {\bibinfo {volume} {11}},\ \bibinfo
  {pages} {041018} (\bibinfo {year} {2021})}\BibitemShut {NoStop}%
\bibitem [{\citenamefont {Brange}\ \emph {et~al.}(2023)\citenamefont {Brange},
  \citenamefont {Pyh\"aranta}, \citenamefont {Heinonen}, \citenamefont
  {Brandner},\ and\ \citenamefont {Flindt}}]{Fredrik2023}%
  \BibitemOpen
  \bibfield  {author} {\bibinfo {author} {\bibfnamefont {F.}~\bibnamefont
  {Brange}}, \bibinfo {author} {\bibfnamefont {T.}~\bibnamefont {Pyh\"aranta}},
  \bibinfo {author} {\bibfnamefont {E.}~\bibnamefont {Heinonen}}, \bibinfo
  {author} {\bibfnamefont {K.}~\bibnamefont {Brandner}},\ and\ \bibinfo
  {author} {\bibfnamefont {C.}~\bibnamefont {Flindt}},\ }\href
  {https://doi.org/10.1103/PhysRevA.107.033324} {\bibfield  {journal} {\bibinfo
   {journal} {Phys. Rev. A}\ }\textbf {\bibinfo {volume} {107}},\ \bibinfo
  {pages} {033324} (\bibinfo {year} {2023})}\BibitemShut {NoStop}%
\bibitem [{\citenamefont {Shen}\ \emph {et~al.}(2023)\citenamefont {Shen},
  \citenamefont {Chen}, \citenamefont {Aliyu}, \citenamefont {Qin},
  \citenamefont {Zhong}, \citenamefont {Loh},\ and\ \citenamefont
  {Lee}}]{PhysRevLett.131.080403}%
  \BibitemOpen
  \bibfield  {author} {\bibinfo {author} {\bibfnamefont {R.}~\bibnamefont
  {Shen}}, \bibinfo {author} {\bibfnamefont {T.}~\bibnamefont {Chen}}, \bibinfo
  {author} {\bibfnamefont {M.~M.}\ \bibnamefont {Aliyu}}, \bibinfo {author}
  {\bibfnamefont {F.}~\bibnamefont {Qin}}, \bibinfo {author} {\bibfnamefont
  {Y.}~\bibnamefont {Zhong}}, \bibinfo {author} {\bibfnamefont
  {H.}~\bibnamefont {Loh}},\ and\ \bibinfo {author} {\bibfnamefont {C.~H.}\
  \bibnamefont {Lee}},\ }\href {https://doi.org/10.1103/PhysRevLett.131.080403}
  {\bibfield  {journal} {\bibinfo  {journal} {Phys. Rev. Lett.}\ }\textbf
  {\bibinfo {volume} {131}},\ \bibinfo {pages} {080403} (\bibinfo {year}
  {2023})}\BibitemShut {NoStop}%
\bibitem [{\citenamefont {Bardeen}\ \emph {et~al.}(1957)\citenamefont
  {Bardeen}, \citenamefont {Cooper},\ and\ \citenamefont
  {Schrieffer}}]{Bardeen:1957tx}%
  \BibitemOpen
  \bibfield  {author} {\bibinfo {author} {\bibfnamefont {J.}~\bibnamefont
  {Bardeen}}, \bibinfo {author} {\bibfnamefont {L.~N.}\ \bibnamefont
  {Cooper}},\ and\ \bibinfo {author} {\bibfnamefont {J.~R.}\ \bibnamefont
  {Schrieffer}},\ }\href {https://doi.org/10.1103/PhysRev.108.1175} {\bibfield
  {journal} {\bibinfo  {journal} {Phys. Rev.}\ }\textbf {\bibinfo {volume}
  {108}},\ \bibinfo {pages} {1175} (\bibinfo {year} {1957})}\BibitemShut
  {NoStop}%
\bibitem [{\citenamefont {Yamamoto}\ \emph {et~al.}(2019)\citenamefont
  {Yamamoto}, \citenamefont {Nakagawa}, \citenamefont {Adachi}, \citenamefont
  {Takasan}, \citenamefont {Ueda},\ and\ \citenamefont
  {Kawakami}}]{Yamamoto2019}%
  \BibitemOpen
  \bibfield  {author} {\bibinfo {author} {\bibfnamefont {K.}~\bibnamefont
  {Yamamoto}}, \bibinfo {author} {\bibfnamefont {M.}~\bibnamefont {Nakagawa}},
  \bibinfo {author} {\bibfnamefont {K.}~\bibnamefont {Adachi}}, \bibinfo
  {author} {\bibfnamefont {K.}~\bibnamefont {Takasan}}, \bibinfo {author}
  {\bibfnamefont {M.}~\bibnamefont {Ueda}},\ and\ \bibinfo {author}
  {\bibfnamefont {N.}~\bibnamefont {Kawakami}},\ }\href
  {https://doi.org/10.1103/PhysRevLett.123.123601} {\bibfield  {journal}
  {\bibinfo  {journal} {Phys. Rev. Lett.}\ }\textbf {\bibinfo {volume} {123}},\
  \bibinfo {pages} {123601} (\bibinfo {year} {2019})}\BibitemShut {NoStop}%
\bibitem [{Note1()}]{Note1}%
  \BibitemOpen
  \bibinfo {note} {Note that in our definition, $U_{R}>0\protect \:(<0)$
  represents attractive (repulsive) interaction.}\BibitemShut {Stop}%
\bibitem [{Sup()}]{SupplementaryMaterial}%
  \BibitemOpen
  \href@noop {} {\bibinfo {title} {See supplemental material for
  details}}\BibitemShut {NoStop}%
\bibitem [{\citenamefont {Coleman}(2015)}]{Coleman:2015vz}%
  \BibitemOpen
  \bibfield  {author} {\bibinfo {author} {\bibfnamefont {P.}~\bibnamefont
  {Coleman}},\ }\href {https://doi.org/DOI: 10.1017/CBO9781139020916} {\emph
  {\bibinfo {title} {Introduction to Many-Body Physics}}}\ (\bibinfo
  {publisher} {Cambridge University Press},\ \bibinfo {address} {Cambridge,
  England},\ \bibinfo {year} {2015})\BibitemShut {NoStop}%
\bibitem [{Note2()}]{Note2}%
  \BibitemOpen
  \bibinfo {note} {In general we can write the Ginzberg-Landau free energy in
  terms of order parameter $\Delta _0$ as $F=a\Delta _0^2+b\Delta _0^4$. Near
  the origin we have $\Delta _0\to 0$ and hence $F\propto \Delta _0^2\propto
  \chi ^2$ holds.}\BibitemShut {Stop}%
\bibitem [{\citenamefont {Shankar}(1994)}]{Shankar1994}%
  \BibitemOpen
  \bibfield  {author} {\bibinfo {author} {\bibfnamefont {R.}~\bibnamefont
  {Shankar}},\ }\href {https://doi.org/10.1103/RevModPhys.66.129} {\bibfield
  {journal} {\bibinfo  {journal} {Rev. Mod. Phys.}\ }\textbf {\bibinfo {volume}
  {66}},\ \bibinfo {pages} {129} (\bibinfo {year} {1994})}\BibitemShut
  {NoStop}%
\bibitem [{\citenamefont {Nakagawa}\ \emph {et~al.}(2018)\citenamefont
  {Nakagawa}, \citenamefont {Kawakami},\ and\ \citenamefont
  {Ueda}}]{Nakagawa2018}%
  \BibitemOpen
  \bibfield  {author} {\bibinfo {author} {\bibfnamefont {M.}~\bibnamefont
  {Nakagawa}}, \bibinfo {author} {\bibfnamefont {N.}~\bibnamefont {Kawakami}},\
  and\ \bibinfo {author} {\bibfnamefont {M.}~\bibnamefont {Ueda}},\ }\href
  {https://doi.org/10.1103/PhysRevLett.121.203001} {\bibfield  {journal}
  {\bibinfo  {journal} {Phys. Rev. Lett.}\ }\textbf {\bibinfo {volume} {121}},\
  \bibinfo {pages} {203001} (\bibinfo {year} {2018})}\BibitemShut {NoStop}%
\bibitem [{\citenamefont {Gersch}\ \emph {et~al.}(2005)\citenamefont {Gersch},
  \citenamefont {Honerkamp}, \citenamefont {Rohe},\ and\ \citenamefont
  {Metzner}}]{Gersch2005}%
  \BibitemOpen
  \bibfield  {author} {\bibinfo {author} {\bibfnamefont {R.}~\bibnamefont
  {Gersch}}, \bibinfo {author} {\bibfnamefont {C.}~\bibnamefont {Honerkamp}},
  \bibinfo {author} {\bibfnamefont {D.}~\bibnamefont {Rohe}},\ and\ \bibinfo
  {author} {\bibfnamefont {W.}~\bibnamefont {Metzner}},\ }\href
  {https://doi.org/10.1140/epjb/e2005-00416-8} {\bibfield  {journal} {\bibinfo
  {journal} {Eur. Phys. J. B}\ }\textbf {\bibinfo {volume} {48}},\ \bibinfo
  {pages} {349} (\bibinfo {year} {2005})}\BibitemShut {NoStop}%
\bibitem [{\citenamefont {Dash}\ and\ \citenamefont
  {S\'en\'echal}(2021)}]{PhysRevB.103.045142}%
  \BibitemOpen
  \bibfield  {author} {\bibinfo {author} {\bibfnamefont {S.~S.}\ \bibnamefont
  {Dash}}\ and\ \bibinfo {author} {\bibfnamefont {D.}~\bibnamefont
  {S\'en\'echal}},\ }\href {https://doi.org/10.1103/PhysRevB.103.045142}
  {\bibfield  {journal} {\bibinfo  {journal} {Phys. Rev. B}\ }\textbf {\bibinfo
  {volume} {103}},\ \bibinfo {pages} {045142} (\bibinfo {year}
  {2021})}\BibitemShut {NoStop}%
\bibitem [{\citenamefont {Grigoriev}\ and\ \citenamefont
  {Lyubshin}(2005)}]{PhysRevB.72.195106}%
  \BibitemOpen
  \bibfield  {author} {\bibinfo {author} {\bibfnamefont {P.~D.}\ \bibnamefont
  {Grigoriev}}\ and\ \bibinfo {author} {\bibfnamefont {D.~S.}\ \bibnamefont
  {Lyubshin}},\ }\href {https://doi.org/10.1103/PhysRevB.72.195106} {\bibfield
  {journal} {\bibinfo  {journal} {Phys. Rev. B}\ }\textbf {\bibinfo {volume}
  {72}},\ \bibinfo {pages} {195106} (\bibinfo {year} {2005})}\BibitemShut
  {NoStop}%
\bibitem [{\citenamefont {Sachdev}(2011)}]{Sachdev:2011uj}%
  \BibitemOpen
  \bibfield  {author} {\bibinfo {author} {\bibfnamefont {S.}~\bibnamefont
  {Sachdev}},\ }\href {https://doi.org/DOI: 10.1017/CBO9780511973765} {\emph
  {\bibinfo {title} {Quantum Phase Transitions}}},\ \bibinfo {edition} {2nd}\
  ed.\ (\bibinfo  {publisher} {Cambridge University Press},\ \bibinfo {address}
  {Cambridge, England},\ \bibinfo {year} {2011})\BibitemShut {NoStop}%
\bibitem [{\citenamefont {Iskin}(2021)}]{PhysRevA.103.013724}%
  \BibitemOpen
  \bibfield  {author} {\bibinfo {author} {\bibfnamefont {M.}~\bibnamefont
  {Iskin}},\ }\href {https://doi.org/10.1103/PhysRevA.103.013724} {\bibfield
  {journal} {\bibinfo  {journal} {Phys. Rev. A}\ }\textbf {\bibinfo {volume}
  {103}},\ \bibinfo {pages} {013724} (\bibinfo {year} {2021})}\BibitemShut
  {NoStop}%
\bibitem [{\citenamefont {Pagano}\ \emph {et~al.}(2015)\citenamefont {Pagano},
  \citenamefont {Mancini}, \citenamefont {Cappellini}, \citenamefont {Livi},
  \citenamefont {Sias}, \citenamefont {Catani}, \citenamefont {Inguscio},\ and\
  \citenamefont {Fallani}}]{PhysRevLett.115.265301}%
  \BibitemOpen
  \bibfield  {author} {\bibinfo {author} {\bibfnamefont {G.}~\bibnamefont
  {Pagano}}, \bibinfo {author} {\bibfnamefont {M.}~\bibnamefont {Mancini}},
  \bibinfo {author} {\bibfnamefont {G.}~\bibnamefont {Cappellini}}, \bibinfo
  {author} {\bibfnamefont {L.}~\bibnamefont {Livi}}, \bibinfo {author}
  {\bibfnamefont {C.}~\bibnamefont {Sias}}, \bibinfo {author} {\bibfnamefont
  {J.}~\bibnamefont {Catani}}, \bibinfo {author} {\bibfnamefont
  {M.}~\bibnamefont {Inguscio}},\ and\ \bibinfo {author} {\bibfnamefont
  {L.}~\bibnamefont {Fallani}},\ }\href
  {https://doi.org/10.1103/PhysRevLett.115.265301} {\bibfield  {journal}
  {\bibinfo  {journal} {Phys. Rev. Lett.}\ }\textbf {\bibinfo {volume} {115}},\
  \bibinfo {pages} {265301} (\bibinfo {year} {2015})}\BibitemShut {NoStop}%
\bibitem [{\citenamefont {H\"ofer}\ \emph {et~al.}(2015)\citenamefont
  {H\"ofer}, \citenamefont {Riegger}, \citenamefont {Scazza}, \citenamefont
  {Hofrichter}, \citenamefont {Fernandes}, \citenamefont {Parish},
  \citenamefont {Levinsen}, \citenamefont {Bloch},\ and\ \citenamefont
  {F\"olling}}]{PhysRevLett.115.265302}%
  \BibitemOpen
  \bibfield  {author} {\bibinfo {author} {\bibfnamefont {M.}~\bibnamefont
  {H\"ofer}}, \bibinfo {author} {\bibfnamefont {L.}~\bibnamefont {Riegger}},
  \bibinfo {author} {\bibfnamefont {F.}~\bibnamefont {Scazza}}, \bibinfo
  {author} {\bibfnamefont {C.}~\bibnamefont {Hofrichter}}, \bibinfo {author}
  {\bibfnamefont {D.~R.}\ \bibnamefont {Fernandes}}, \bibinfo {author}
  {\bibfnamefont {M.~M.}\ \bibnamefont {Parish}}, \bibinfo {author}
  {\bibfnamefont {J.}~\bibnamefont {Levinsen}}, \bibinfo {author}
  {\bibfnamefont {I.}~\bibnamefont {Bloch}},\ and\ \bibinfo {author}
  {\bibfnamefont {S.}~\bibnamefont {F\"olling}},\ }\href
  {https://doi.org/10.1103/PhysRevLett.115.265302} {\bibfield  {journal}
  {\bibinfo  {journal} {Phys. Rev. Lett.}\ }\textbf {\bibinfo {volume} {115}},\
  \bibinfo {pages} {265302} (\bibinfo {year} {2015})}\BibitemShut {NoStop}%
\bibitem [{\citenamefont {Zhang}\ \emph {et~al.}(2015)\citenamefont {Zhang},
  \citenamefont {Cheng}, \citenamefont {Zhai},\ and\ \citenamefont
  {Zhang}}]{Zhang2015}%
  \BibitemOpen
  \bibfield  {author} {\bibinfo {author} {\bibfnamefont {R.}~\bibnamefont
  {Zhang}}, \bibinfo {author} {\bibfnamefont {Y.}~\bibnamefont {Cheng}},
  \bibinfo {author} {\bibfnamefont {H.}~\bibnamefont {Zhai}},\ and\ \bibinfo
  {author} {\bibfnamefont {P.}~\bibnamefont {Zhang}},\ }\href
  {https://doi.org/10.1103/PhysRevLett.115.135301} {\bibfield  {journal}
  {\bibinfo  {journal} {Phys. Rev. Lett.}\ }\textbf {\bibinfo {volume} {115}},\
  \bibinfo {pages} {135301} (\bibinfo {year} {2015})}\BibitemShut {NoStop}%
\bibitem [{\citenamefont {Tomita}\ \emph {et~al.}(2017)\citenamefont {Tomita},
  \citenamefont {Nakajima}, \citenamefont {Danshita}, \citenamefont {Takasu},\
  and\ \citenamefont {Takahashi}}]{doi:10.1126/sciadv.1701513}%
  \BibitemOpen
  \bibfield  {author} {\bibinfo {author} {\bibfnamefont {T.}~\bibnamefont
  {Tomita}}, \bibinfo {author} {\bibfnamefont {S.}~\bibnamefont {Nakajima}},
  \bibinfo {author} {\bibfnamefont {I.}~\bibnamefont {Danshita}}, \bibinfo
  {author} {\bibfnamefont {Y.}~\bibnamefont {Takasu}},\ and\ \bibinfo {author}
  {\bibfnamefont {Y.}~\bibnamefont {Takahashi}},\ }\href
  {https://doi.org/10.1126/sciadv.1701513} {\bibfield  {journal} {\bibinfo
  {journal} {Sci. Adv.}\ }\textbf {\bibinfo {volume} {3}},\ \bibinfo {pages}
  {e1701513} (\bibinfo {year} {2017})}\BibitemShut {NoStop}%
\bibitem [{\citenamefont {Honda}\ \emph {et~al.}(2023)\citenamefont {Honda},
  \citenamefont {Taie}, \citenamefont {Takasu}, \citenamefont {Nishizawa},
  \citenamefont {Nakagawa},\ and\ \citenamefont {Takahashi}}]{Honda2023}%
  \BibitemOpen
  \bibfield  {author} {\bibinfo {author} {\bibfnamefont {K.}~\bibnamefont
  {Honda}}, \bibinfo {author} {\bibfnamefont {S.}~\bibnamefont {Taie}},
  \bibinfo {author} {\bibfnamefont {Y.}~\bibnamefont {Takasu}}, \bibinfo
  {author} {\bibfnamefont {N.}~\bibnamefont {Nishizawa}}, \bibinfo {author}
  {\bibfnamefont {M.}~\bibnamefont {Nakagawa}},\ and\ \bibinfo {author}
  {\bibfnamefont {Y.}~\bibnamefont {Takahashi}},\ }\href
  {https://doi.org/10.1103/PhysRevLett.130.063001} {\bibfield  {journal}
  {\bibinfo  {journal} {Phys. Rev. Lett.}\ }\textbf {\bibinfo {volume} {130}},\
  \bibinfo {pages} {063001} (\bibinfo {year} {2023})}\BibitemShut {NoStop}%
\bibitem [{\citenamefont {Nakagawa}\ \emph {et~al.}(2021)\citenamefont
  {Nakagawa}, \citenamefont {Kawakami},\ and\ \citenamefont
  {Ueda}}]{PhysRevLett.126.110404}%
  \BibitemOpen
  \bibfield  {author} {\bibinfo {author} {\bibfnamefont {M.}~\bibnamefont
  {Nakagawa}}, \bibinfo {author} {\bibfnamefont {N.}~\bibnamefont {Kawakami}},\
  and\ \bibinfo {author} {\bibfnamefont {M.}~\bibnamefont {Ueda}},\ }\href
  {https://doi.org/10.1103/PhysRevLett.126.110404} {\bibfield  {journal}
  {\bibinfo  {journal} {Phys. Rev. Lett.}\ }\textbf {\bibinfo {volume} {126}},\
  \bibinfo {pages} {110404} (\bibinfo {year} {2021})}\BibitemShut {NoStop}%
\bibitem [{\citenamefont {Ashida}\ \emph {et~al.}(2016)\citenamefont {Ashida},
  \citenamefont {Furukawa},\ and\ \citenamefont {Ueda}}]{PhysRevA.94.053615}%
  \BibitemOpen
  \bibfield  {author} {\bibinfo {author} {\bibfnamefont {Y.}~\bibnamefont
  {Ashida}}, \bibinfo {author} {\bibfnamefont {S.}~\bibnamefont {Furukawa}},\
  and\ \bibinfo {author} {\bibfnamefont {M.}~\bibnamefont {Ueda}},\ }\href
  {https://doi.org/10.1103/PhysRevA.94.053615} {\bibfield  {journal} {\bibinfo
  {journal} {Phys. Rev. A}\ }\textbf {\bibinfo {volume} {94}},\ \bibinfo
  {pages} {053615} (\bibinfo {year} {2016})}\BibitemShut {NoStop}%
\bibitem [{Note3()}]{Note3}%
  \BibitemOpen
  \bibinfo {note} {To be precise, the gap equation at finite temperature is
  singular on the critical line since the Yang-Lee zeros make the expectation
  value ill-defined. However, we can define the value of the partition function
  on the critical line from the continuity of the partition function in a
  finite-size system.}\BibitemShut {Stop}%
\bibitem [{\citenamefont {Kato}(1995)}]{kato1995perturbation}%
  \BibitemOpen
  \bibfield  {author} {\bibinfo {author} {\bibfnamefont {T.}~\bibnamefont
  {Kato}},\ }\href {https://doi.org/10.1007/978-3-642-66282-9} {\emph {\bibinfo
  {title} {Perturbation Theory for Linear Operators}}},\ \bibinfo {edition}
  {2nd}\ ed.\ (\bibinfo  {publisher} {Springer Berlin, Heidelberg},\ \bibinfo
  {year} {1995})\BibitemShut {NoStop}%
\end{thebibliography}%

\clearpage{}

\onecolumngrid
\appendix
\renewcommand{\thefigure}{S\arabic{figure}}
\setcounter{figure}{0} 
\renewcommand{\thepage}{S\arabic{page}}
\setcounter{page}{1} 
\renewcommand{\theequation}{S.\arabic{equation}}
\setcounter{equation}{0} 
\renewcommand{\thesection}{S\arabic{section}}
\setcounter{section}{0}

\begin{center}
	\large{Supplemental Material for}\\
	\textbf{``Yang-Lee Zeros, Semicircle Theorem, and Nonunitary Criticality in Bardeen-Cooper-Schrieffer Superconductivity"}
\end{center}

\tableofcontents{}
\section{Yang-Lee Zeros on the Phase Boundary}
We begin from considering the three-dimensional non-Hermitian BCS Hamiltonian
\begin{equation}
	H=\sum_{\boldsymbol{k}\sigma}\xi_{\boldsymbol{k}}c_{\boldsymbol{k}\sigma}^{\dagger}c_{\boldsymbol{k}\sigma}-\frac{U}{N}\sum_{\bm{k},\bm{k}'}{}^{'}c_{\bm{k}\uparrow}^{\dagger}c_{\bm{-k}\downarrow}^{\dagger}c_{\bm{-k}'\downarrow}c_{\bm{k}'\uparrow}, 
\end{equation}
where $U=U_R+iU_I$ and the prime in $\sum_{\bm{k}}^{'}$ indicates that the sum over $\bm{k}$ restricted to  $|\xi_{\boldsymbol{k}}|<\omega_D$ with $\omega_D$ being the cutoff energy. The mean-field Hamiltonian is given by 
\begin{equation}
	H_{\mathrm{MF}}=\sum_{\boldsymbol{k}\sigma}\xi_{\boldsymbol{k}}c_{\boldsymbol{k}\sigma}^{\dagger}c_{\boldsymbol{k}\sigma}+\sum_{\bm{k}}{}^{'}[\bar{\Delta}_0c_{-\bm{k}\downarrow}c_{\bm{k}\uparrow}+\Delta_0 c_{\bm{k}\uparrow}^{\dagger}c_{-\bm{k}\downarrow}^{\dagger}]+\frac{N}{U}\bar{\Delta}_{0}\Delta_{0},
\end{equation}
where $\Delta_{0}=-\frac{U}{N}\sum_{\boldsymbol{k}L}\langle c_{-\boldsymbol{k}\downarrow}c_{\boldsymbol{k}\uparrow}\rangle_{\mathrm{R}}$ and $\bar{\Delta}_0=-\frac{U}{N}\sum_{\boldsymbol{k}L}\langle c^{\dagger}_{\boldsymbol{k}\uparrow}c^{\dagger}_{-\boldsymbol{k}\downarrow}\rangle_{\mathrm{R}}$ represent the superconducting gap. The right and left ground states of the mean-filed Hamiltonian $H_{\mathrm{MF}}$ are given by \cite{Yamamoto2019}
\begin{align}
	|\text{BCS}\rangle_{R}&=\prod_{\bm{k}}(u_{\bm{k}}+v_{\bm{k}}c_{\boldsymbol{k}\uparrow}^{\dagger}c_{-\boldsymbol{k}\downarrow}^{\dagger})|0\rangle,\\
	|\text{BCS}\rangle_{L}&=\prod_{\bm{k}}(u^{*}_{\bm{k}}+\bar{v}^{*}_{\bm{k}}c_{\boldsymbol{k}\uparrow}^{\dagger}c_{-\boldsymbol{k}\downarrow}^{\dagger})|0\rangle,
\end{align}
where the parameters $u_{\bm{k}},v_{\bm{k}}$ and $\bar{v}_{\bm{k}}$ are complex coefficients and take the specific form of
\begin{equation}
	u_{\bm{k}}=\sqrt{\frac{E_{\bm{k}}+\xi_{\bm{k}}}{2E_{\bm{k}}}},\quad v_{\bm{k}}=-\sqrt{\frac{E_{\bm{k}}-\xi_{\bm{k}}}{2E_{\bm{k}}}}\sqrt{\frac{\Delta_0}{\bar{\Delta}_0}},\quad
	\bar{v}_{\bm{k}}=-\sqrt{\frac{E_{\bm{k}}-\xi_{\bm{k}}}{2E_{\bm{k}}}}\sqrt{\frac{\bar{\Delta}_0}{\Delta_0}}.
\end{equation}
Here $E_{\bm{k}}=\sqrt{\xi_{\bm{k}}^2+\Delta_{\bm{k}}^2}$ is the dispersion relation of Bogoliubov quasiparticles where $\Delta_{\bm{k}}=\Delta_0\theta(\omega_D-|\xi_{\bm{k}}|)$ with $\theta(x)$ being the Heaviside step function. Note that $\bar{\Delta}_0\neq\Delta_0^*$. In the following we take a gauge  \cite{Yamamoto2019} in which $\Delta_0=\bar{\Delta}_0\in\mathbb{C}$.

The gap equation at absolute zero reads as \cite{Yamamoto2019}
\begin{equation}
	\frac{N}{U}=\sum_{\bm{k}}{}^{'}\frac{1}{2\sqrt{\xi_{\bm{k}}^2+\Delta_0^2}}.
\end{equation}
Provided that the density of states is constant and given by $\rho_0$, the above equation can be simplified as
\begin{equation}
	\frac{\sqrt{\omega_D^2 + \Delta_0^2} + \omega_D}{\Delta_0} =
	e^{\frac{1}{\rho_0 U}}\,.
\end{equation}
The solution to the gap equation is given by $\Delta_0 =
\frac{\omega_D}{\text{sinh} \left( \frac{1}{\rho_0 U} \right)}$. To be specific,
\begin{eqnarray}
	\Delta_0 & = & \frac{2 \omega_D}{\text{exp} \left[ \frac{1}{\rho_0 | U |^2}
		\left( U_R - i U_I \right) \right] - \text{exp} \left[ -
		\frac{1}{\rho_0 | U |^2} \left( U_R - iU_I \right) \right]}
	\nonumber\\
	& = & \frac{\omega_D}{\text{sinh} \left( \frac{U_R}{\rho_0 | U |^2} \right)
		\cos \left( \frac{U_I}{\rho_0 | U |^2} \right) - i \text{cosh}
		\left( \frac{U_R}{\rho_0 | U |^2} \right) \sin \left( \frac{U_I}{\rho_0 | U |^2} \right)}, 
	\label{Delta_0}
\end{eqnarray}
and its real part is given by
\begin{equation}
	\text{Re} [\Delta_0] = \omega_D \frac{\text{sinh} \left( \frac{U_R}{\rho_0 |
			U |^2} \right) \cos \left( \frac{U_I}{ \rho_0 | U |^2}
		\right)}{\left( \text{sinh} \left( \frac{U_R}{\rho_0 | U |^2} \right)
		\cos \left( \frac{U_I}{ \rho_0 | U |^2} \right) \right)^2 + \left(
		\text{cosh} \left( \frac{U_R}{\rho_0 | U |^2} \right) \sin \left(
		\frac{U_I}{ \rho_0 | U |^2} \right) \right)^2}\,.
	\label{RDelta_0}
\end{equation}
At the quantum phase transition point, the real part of the gap vanishes, which gives $\cos \left( \frac{U_I}{ \rho_0 | U |^2} \right) = 0$, or equivalently, $\frac{U_I}{ \rho_0 | U |^2} = \frac{\pi}{2}$. This determines the condition for the phase boundary \cite{Yamamoto2019}
\begin{equation}
	(\rho_0 \pi U_R)^2 + (\rho_0 \pi U_I- 1)^2 = 1
	\label{phase_transition}\,.
\end{equation}
This condition restricts the imaginary part of the gap $\Delta_0$ as
\begin{eqnarray}
	\text{Im} [\Delta_0] & = & \omega_D \frac{\text{cosh} \left(
		\frac{U_R}{\rho_0 | U |^2} \right) \sin \left( \frac{U_I}{ \rho_0
			| U |^2} \right)}{\left( \text{sinh} \left( \frac{U_R}{\rho_0 | U |^2}
		\right) \cos \left( \frac{U_I}{ \rho_0 | U |^2} \right) \right)^2
		+ \left( \text{cosh} \left( \frac{U_R}{\rho_0 | U |^2} \right) \sin
		\left( \frac{U_I}{ \rho_0 | U |^2} \right) \right)^2} \nonumber\\
	& = & \frac{\omega_D}{\text{cosh} \left( \frac{U_R}{\rho_0 | U |^2}
		\right)}\,.
	\label{IDelta_0}
\end{eqnarray}

The phase transition is related to the existence of Yang-Lee zeros on the phase boundary. The partition function of Bogoliubov quasi-particles in a finite-size system at finite temperature $1/\beta$ is
\begin{equation}
	Z = \prod_{\bm{k},\sigma} (1 + e^{-\beta E_{\bm{k}}})\,.
\end{equation}
Since we only consider the case with a large $\beta$, we directly substitute Eq. (\ref{Delta_0}) into the dispersion relation \footnote{To be precise, the gap equation at finite temperature is singular on the critical line since the Yang-Lee zeros make the expectation value ill-defined. However, we can define the value of the partition function on the critical line from the continuity of the partition function in a finite-size system.}. For the points not on the phase boundary, we have $\text{Re}
[E_{\bm{k}}] > 0$, which indicates that the
partition function cannot vanish. However, on the critical line (\ref{phase_transition}), the gap $\Delta_0$ becomes purely imaginary and therefore Yang-Lee zeros can emerge. The partition function on the phase boundary can be decomposed as
\begin{equation}
	Z = \prod_{\bm{k},|\xi_{\bm{k}}|<\text{Im}\Delta_0,\sigma} \left( 1 + e^{- i \beta \sqrt{\left( \text{Im}\Delta_0 \right)^2 - \xi_{\bm{k}}^2}} \right)\times \prod_{\bm{k},|\xi_{\bm{k}}|>\text{Im}\Delta_0,\sigma}\left( 1 + e^{- \beta \sqrt{\xi_{\bm{k}}^2 +\Delta_{\bm{k}}^2 }} \right).
\end{equation}
The first product vanishes for the momentum $\bm{k}$ that satisfies the condition
\begin{equation}
	\beta \sqrt{\left( \text{Im} \Delta_0 \right)^2 - \xi_{\bm{k}}^2} = (2n + 1) \pi,
\end{equation}
where $n$ is an arbitrary integer. This condition is equivalent to
\begin{equation}
	|\xi_{\bm{k}}| = \sqrt{\left( \text{Im} \Delta_0
		\right)^2 - \left( \frac{2 n + 1}{\beta} \pi \right)^2}.
	\label{condition}
\end{equation}
Further, we take the thermodynamic limit. Since $\text{Im}\Delta_0<\omega_D$, we can always find the momentum $\bm{k}$ in the energy shell satisfying the condition (\ref{condition}) for an arbitrarily large $\beta$. Hence, Yang-Lee zeros are distributed on the phase boundary (\ref{phase_transition}). 

\section{Renormalization Group Theory of Non-Hermitian BCS Superconductivity}\label{RG}

Here we consider the renormalization-group (RG) flow of the interaction strength to elucidate that the Yang-Lee singularity corresponds to the RG critical line. The one-loop beta function $\beta_1(U)$ is given at the order of $U^2$ by \cite{Shankar1994}
\begin{equation}
	\frac{dV}{dt}=V^{2}=:\beta_1(U)\,,
	\label{one-loop}
\end{equation}
where $dt=-\frac{d\Xi}{\Xi}$ is the relative width of the high-energy
shell, $\Xi$ is the cutoff of the energy $\xi_{\bm{k}}$ and $V=\rho_0 U$ is the dimensionless interaction strength. Here $t$ is considered as the RG-flow parameter. We take the two-loop correction into account and consider the terms of the order of $U^3$. After the two-loop calculation, we will see that the RG equation reproduces the phase boundary shown in Fig. 1 in the main text.

\begin{figure}{\tiny }
	\includegraphics[width=0.4\columnwidth]{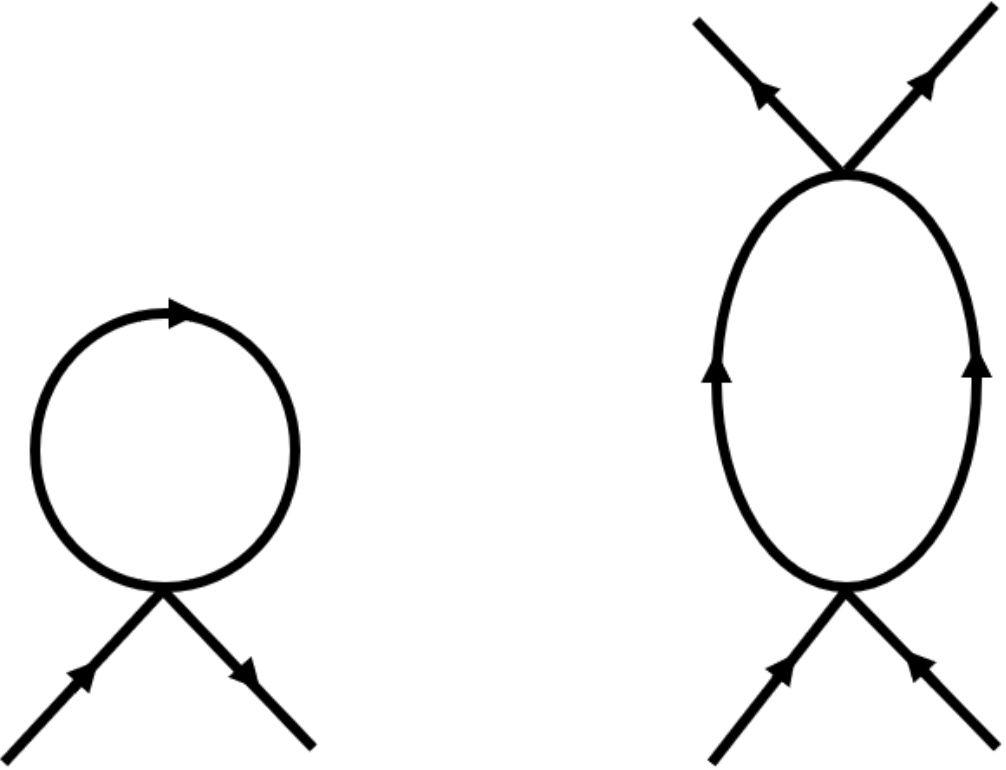}
	\caption{Feynman diagrams for the self-energy correction and the renormalization of the interaction strength. The right diagram is the BCS diagram.}
	
	\label{feynman_diagram1}
\end{figure}

\begin{figure}
	\includegraphics[width=1\columnwidth]{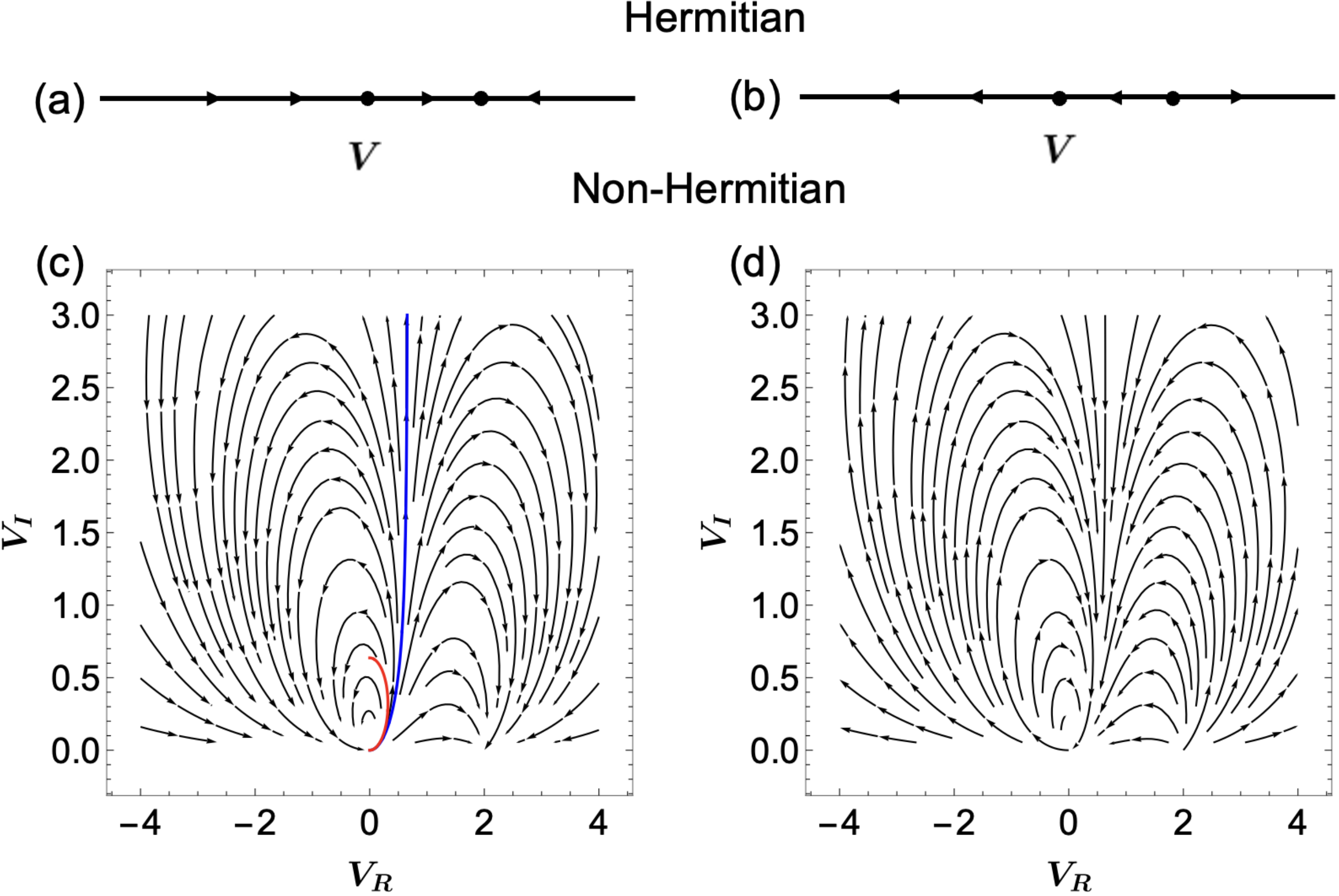}
	\caption{Canonical RG flow for coupling strength $V$ for general Hermitian and non-Hermitian Hamiltonians with the RG flow equation $\frac{dV}{dt}=aV^2+bV^3$. In the diagram (a) and (c), the parameters are set to be $a=1$ and $b=-1/2$. In the diagram (b) and (d), the parameters are set to be $a=-1$ and $b=1/2$. For $b>0$, there is neither a nontrivial stable fixed point nor a critical line.}
	
	\label{RG_Flow_Diagram}
\end{figure}

Up to just one integral over the momenta, the higher-order contribution
to the beta function comes from the correction for the high-momentum propagator.
Actually, the self-energy for the high-momentum
propagator shown in Fig. \ref{feynman_diagram1} is given by

\begin{equation}
	\Sigma(\bm{k},\Omega)=\int\frac{d\omega}{2\pi}\frac{U}{i\omega-\xi_{\bm{k}}}e^{i\omega0^+}=U\theta(-\xi_{\bm{k}}),
\end{equation}
where $\Omega$ is the frequency for the external leg. Here we introduce a factor $e^{i\omega0^+}$ to ensure the convergence \cite{Shankar1994}. Therefore, the propagator is modified as

\begin{equation}
	G(\bm{k},\Omega)=\frac{1}{i\Omega-\xi_{\bm{k}}}+\frac{U\theta(-\xi_{\bm{k}})}{(i\Omega-\xi_{\bm{k}})^{2}}\,.
\end{equation}
By including the self-energy diagram in Fig. \ref{feynman_diagram1}, we can find the corrected contribution from the BCS diagram. After integrating out the energy shell $(-\Xi,-\Xi+d\Xi)$ of $\xi_{\bm{k}}$, we obtain the two-loop correction $\beta_2(U)$ to the beta function as
\begin{align}
	\beta_2(U)&= \frac{1}{2}\rho_0\Xi U^{2}\left(\int\frac{d\Omega}{2\pi}\frac{U}{(i\Omega-\Xi)^{2}}\frac{1}{-i\Omega-\Xi}+\int\frac{d\Omega}{2\pi}\frac{1}{i\Omega-\Xi}\frac{U}{(i\Omega+\Xi)^{2}}\right)\\
	& =-\frac{\rho_0^2U^3}{2}\,,
	\label{two-loop}
\end{align}
where we define $\rho_0=1/(2\Xi)$ since $\frac{1}{N}\sum_{\bm{k}}=1=\int\rho_0d\xi_{\bm{k}}$ is satisfied \cite{Yamamoto2019}. 
Hence, the RG equation up to two-loop order is written as
\begin{equation}
	\frac{dV}{dt}=V^{2}-\frac{1}{2}V^{3}\,.\label{eq:RG_Flow_Equ_Total}
\end{equation}
The RG flow diagram for Eq. (\ref{eq:RG_Flow_Equ_Total}) is shown in Fig. \ref{RG_Flow_Diagram}(c).

As can be seen from Eq. (\ref{eq:RG_Flow_Equ_Total}), the RG flow has a nontrivial fixed point at $V=2$ and a critical line depicted as the blue curve in Fig. \ref{RG_Flow_Diagram}. This critical line separates the whole space into two phases, with one flowing to the origin and the other flowing to the nontrivial fixed point. The analytical expression for the critical line can be derived as follows \cite{Nakagawa2018}. We rewrite the RG equation (\ref{eq:RG_Flow_Equ_Total}) as

\begin{align}
	\frac{dV_{R}}{dt} & =V_{R}^{2}-V_{I}^{2}-\frac{1}{2}V_{R}^{3}+\frac{3}{2}V_{R}V_{I}^{2},\\
	\frac{dV_{I}}{dt} & =2V_{R}V_{I}-\frac{3}{2}V_{I}V_{R}^{2}+\frac{1}{2}V_{I}^{3},
\end{align}
where $V_{R}=\mathrm{Re}(V),V_{I}=\mathrm{Im}(V)$. On the critical line, the interaction parameter flows towards $V_R=\frac{2}{3},V_I=\infty$, which can be derived from the condition $\frac{dV_R}{dt}=0$ with $V_I\to\infty$.  The specific expression of the critical line can be obtained through integration of Eq. (\ref{eq:RG_Flow_Equ_Total}) as

\begin{equation}
	t=-\frac{1}{V^{\mathrm{f}}}+\frac{1}{2}\ln V^{\mathrm{f}}-\frac{1}{2}\ln(2-V^{\mathrm{f}})+\frac{1}{V}-\frac{1}{2}\ln V+\frac{1}{2}\ln(2-V)\label{eq:Relation_between_t_and_U}\,,
\end{equation}
where the superscript f denotes the final value of the interaction parameter. Since $V_{I}^{\mathrm{f}}\to\infty$ and $V_{R}^{\mathrm{f}}\to\frac{2}{3}$ on the critical line,
the imaginary part of Eq. (\ref{eq:Relation_between_t_and_U}) reads as 
\begin{equation}
	0=\frac{\pi}{2}-\frac{V_{I}}{V_{R}^{2}+V_{I}^{2}}-\frac{1}{2}\arctan\frac{V_{I}}{V_{R}}-\frac{1}{2}\arctan\frac{V_{I}}{2-V_{R}}\,.
	\label{eq:Phase_boundary_by_RG}
\end{equation}
This is the equation defining the phase boundary. Around the origin, the expression of this critical line (\ref{eq:Phase_boundary_by_RG}) can be expanded as
\begin{equation}
	0=-\frac{V_{I}}{V_{R}^{2}+V_{I}^{2}}+\frac{\pi}{2}\,,
\end{equation}
which is consistent with the phase boundary (\ref{phase_transition}) obtained from the mean-field theory. The RG result confirms the validity of the mean-field analysis.

By taking $V^{\mathrm{f}}$ to be pure imaginary, we obtain the energy scale $T_{\mathrm{recur}}$ that characterizes the reversion of the RG flow \cite{Nakagawa2018}:
\begin{equation}
	t=\frac{i}{V_{I}^{\mathrm{f}}}+\frac{1}{2}\ln V_{I}^{\mathrm{f}}+i\frac{\pi}{4}-\frac{1}{4}\ln(4+(V_{I}^{\mathrm{f}})^{2})-\frac{i}{2}(2\pi-\arctan\frac{V_{I}^{\mathrm{f}}}{2})+\frac{1}{V}-\frac{1}{2}\ln V+\frac{1}{2}\ln(2-V)\,.
	\label{Delta t}
\end{equation}
If we assume $|V|\ll1$, the above equation (\ref{Delta t}) can be rewritten as
\begin{align}
	0 & =\frac{1}{V_{I}^{\mathrm{f}}}+\frac{\pi}{4}+\frac{1}{2}\arctan\frac{V_{I}^{\mathrm{f}}}{2}-\frac{V_{I}}{V_{R}^{2}+V_{I}^{2}}-\frac{1}{2}\arctan\frac{V_{I}}{V_{R}}-\frac{1}{2}\arctan(\frac{V_{I}}{2-V_{R}}),\\
	e^{-t} & =\sqrt{\frac{\sqrt{4+(V_{I}^{\mathrm{f}})^{2}}}{V_{I}^{\mathrm{f}}}}\exp(-\frac{V_{R}}{V_{R}^{2}+V_{I}^{2}})(\frac{|V|}{|2-V|})^{\frac{1}{2}}.
\end{align}
From the second equation, we have the reversion temperature $T_\mathrm{recur}$ at which the RG flow reaches a point $(0,V_I^f)$ on the imaginary axis:
\begin{equation}
	T_{\mathrm{recur}}\sim e^{-t}=\sqrt{\frac{\sqrt{4+(V_{I}^{\mathrm{f}})^{2}}}{V_{I}^{\mathrm{f}}}}\exp(-\frac{V_{R}}{V_{R}^{2}+V_{I}^{2}})(\frac{|V|}{|2-V|})^{\frac{1}{2}}\,.
\end{equation}
Near the phase boundary in which $V_I^f\gg1$, we can obtain the simplified expression for the reversion temperature as
\begin{equation}
	T_{\mathrm{recur}}\sim e^{-t}=\exp(-\frac{V_{R}}{V_{R}^{2}+V_{I}^{2}})(\frac{|V|}{|2-V|})^{\frac{1}{2}}\,.
\end{equation}

Finally, we consider a general canonical RG equation for a marginal interaction $V$ up to the order of $V^3$:
\begin{equation}
	\frac{dV}{dt}=aV^2+bV^3.
	\label{eq:RG_Flow_Equ_Marginal_canonical}
\end{equation}
In Fig. \ref{RG_Flow_Diagram}, we show both the Hermitian case and the non-Hermitian case. We find that there are only two types of RG flows for a marginal interaction. The first type with $b<0$ is shown in Fig. \ref{RG_Flow_Diagram}(a,c). In the Hermitian case, it has a nontrivial stable fixed point, whereas in the non-Hermitian case, it has a critical line. The second type with $b>0$ is shown in Fig. \ref{RG_Flow_Diagram}(b,d). In the Hermitian case, it has a non-trivial unstable fixed point, whereas in the non-Hermitian case, it shows no critical line. Hence, we can see  that only the first type of RG flows has a phase boundary and a phase transition.

We can also derive the critical line for the general canonical RG flow with arbitrary $a$ and $b<0$. Integrating Eq. (\ref{eq:RG_Flow_Equ_Marginal_canonical}) with respect to $t$, we obtain
\begin{equation}
	t=-\frac{1}{a V^{\mathrm{f}}}-\frac{b}{a^2}\ln{V^\mathrm{f}}+\frac{b}{a^2}\ln{(a+b V^\mathrm{f})}+\frac{1}{a V}+\frac{b}{a^2}\ln{V}-\frac{b}{a^2}\ln{(a+b V)}.
	\label{eq:general_formula_final}
\end{equation}
On the critical line for $V_I\to\infty$ and $V_R\to-\frac{a}{3b}$, the imaginary part of Eq. (\ref{eq:general_formula_final}) reads as 
\begin{equation}
	0=-\frac{b\pi}{a|a|}-\frac{1}{a}\frac{V_I}{V_R^2+V_I^2}+\frac{b}{a^2}\arctan{\frac{V_I}{V_R}}+\frac{b}{a^2}\arctan{(-\frac{b V_I}{a+b V_R})}.
	\label{eq:imaginary_part_boundary}
\end{equation}
Near the origin, this critical line can be expanded as 
\begin{equation}
	0=-\frac{V_I}{V_R^2+V_I^2}-\frac{b\pi}{|a|}\,,
\end{equation}
where $V_R>0$ if $a>0$ and $V_R<0$ if $a<0$, which are both a semicircle.

\section{Correlation Functions on the Phase Boundary}\label{SectionOn}

\begin{figure}
	\centering \includegraphics[width=0.6\columnwidth]{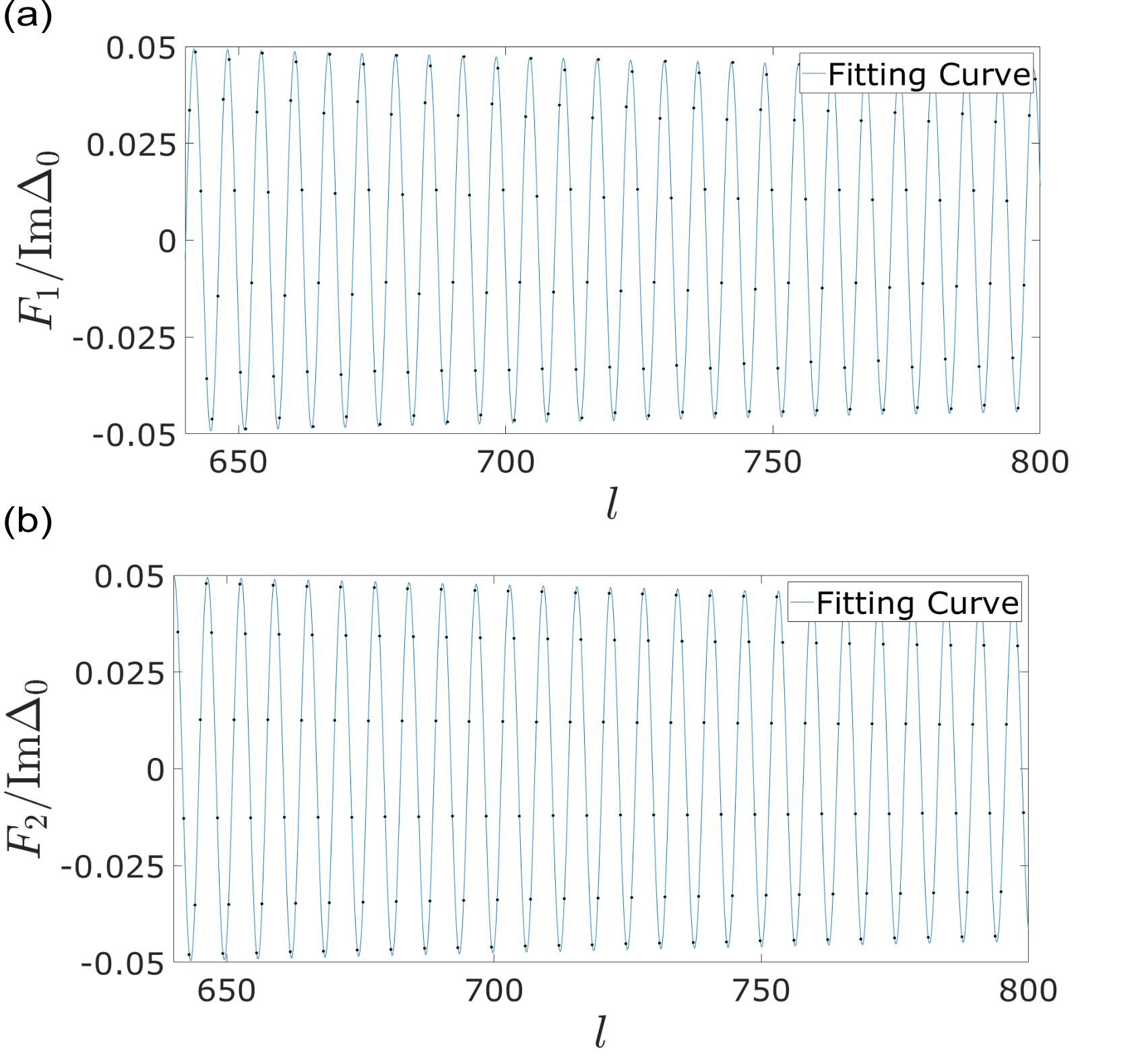}
	\caption{(a) Real part $F_1(l)$ and (b) imaginary part $F_2(l)$ of the correlation function on the phase boundary which are defined in Eqs. (\ref{definition_of_F1}) and (\ref{definition_of_F2}). Here $(\rho_{0}U_{R},\rho_{0}U_{I})=(\frac{1}{\pi},\frac{1}{\pi})$ and $l=\frac{\text{Im}\Delta_0}{v_F}x$. The fitting parameters are shown in Table \ref{fitting_table_2}.}
	\label{fig2}
\end{figure}

In this section, we calculate the correlation functions on the phase boundary to elucidate the critical behavior at the Yang-Lee singularity. Here we firstly consider the momentum distribution of the particles
\begin{equation}
	{}_L\langle c_{\bm{k} \sigma}^{\dagger} c_{\bm{k} \sigma} \rangle_R = {}_L \left\langle
	\text{BCS} \right| c_{\bm{k} \sigma}^{\dagger} c_{\bm{k} \sigma} \left| \text{BCS}\right\rangle_R \,.
\end{equation}
Using the expression of the BCS states, we obtain
\begin{equation}
	{}_L\langle c_{\bm{k} \uparrow}^{\dagger} c_{\bm{k} \uparrow} \rangle_R = {}_L\langle c_{\bm{k}\downarrow}^{\dagger} c_{\bm{k} \downarrow} \rangle_R = v_{\bm{k}}^2 = \frac{1}{2} -
	\frac{\xi_{\bm{k}}}{2 E_{\bm{k}}} = \frac{1}{2} - \frac{\xi_{\bm{k}}}{2 \sqrt{\xi_{\bm{k}}^2 +
			\Delta_{\bm{k}}^2}}\,.
\end{equation}
Similarly, we have 
\begin{equation}
	{}_L\langle c_{\bm{k} \uparrow}^{\dagger} c_{\bm{k} \downarrow} \rangle_R = {}_L\langle c_{\bm{k}\downarrow}^{\dagger} c_{\bm{k} \uparrow} \rangle_R = 0\,.
\end{equation}
Then we perform the Fourier transformation to 
\begin{equation}
	C (\bm{x}-\bm{x}') := {}_L\langle c_{\sigma}^{\dagger} (\bm{x}) c_{\sigma} (\bm{x}') \rangle_R = \int \frac{d^3 \bm{k}}{(2 \pi)^3} \left(\frac{1}{2} - \frac{\xi_{\bm{k}}}{2 \sqrt{\xi_{\bm{k}}^2 + \Delta_{\bm{k}}^2}} \right)e^{i\bm{k} \cdot (\bm{x}-\bm{x}')}\,.
	\label{eq:define_CF}
\end{equation}
We drop the first term on the right-hand side of Eq. (\ref{eq:define_CF}) since it is proportional to the delta function $\delta(\bm{x})$. Here, we replace the integral with $\int ' \frac{d^3 \bm{k}}{(2 \pi)^3}$ for the momentum with $|\xi_{\bm{k}}|<\omega_D$ since we are only concerned with the long-range behavior of the correlation function. 
In the following, we will replace $\bm{x}-\bm{x}'$
with $\bm{x}$ for convenience. Then the correlation
function is given by
\begin{equation}
	C (\bm{x}) \simeq -\int ' \frac{d^3 \bm{k}}{(2 \pi)^3} \frac{\xi_{\bm{k}}}{2 \sqrt{\xi_{\bm{k}}^2
			+ \Delta_0^2}} e^{i\bm{k} \cdot \bm{x}}.
\end{equation}
To extract the long-range behavior of the correlation function, we expand the energy spectrum around the Fermi surface as $\xi_{\bm{k}} \simeq v_F (k - k_F)$, where $v_F$ and $k_F$ are the Fermi velocity and the Fermi momentum, respectively. Then the integration can be simplified as
\begin{equation}
	C(\bm{x})\simeq-\int ' \frac{\rho_0}{2} d \xi_{\bm{k}} \sin \theta d \theta \frac{\xi_{\bm{k}}}{2
		\sqrt{\xi_{\bm{k}}^2 + \Delta_0^2}} e^{i \frac{\xi_{\bm{k}}}{v_F} x \cos \theta}
	e^{i k_F x \cos \theta}\,,
\end{equation}
where $x=|\bm{x}|$ and $\rho_0$ is the density of states at the Fermi surface. With the
integration over $\theta$, we have
\begin{align}
	C(\bm{x})&\simeq-\frac{1}{2} \int_{- \omega_D}^{\omega_D} \rho_0 d \xi_{\bm{k}} \frac{\xi_{\bm{k}}
		\sin \left( \left( \frac{\xi_{\bm{k}}}{v_F} + k_F \right) x \right)}{\left(
		\frac{\xi_{\bm{k}}}{v_F} + k_F \right) x \sqrt{\xi_{\bm{k}}^2 + \Delta_0^2}}\nonumber\\
	& =  -\rho_0 v_F \int_0^{\omega_D} d \xi_{\bm{k}} \frac{\xi_{\bm{k}}}{x \sqrt{\xi_{\bm{k}}^2 +
			\Delta_0^2}} \left[ (- \xi_{\bm{k}}) \frac{\sin (k_F x) \cos \left(
		\frac{\xi_{\bm{k}}}{v_F} x \right)}{(v_F k_F)^2 - \xi_{\bm{k}}^2} + v_F k_F
	\frac{\cos (k_F x) \sin \left( \frac{\xi_{\bm{k}}}{v_F} x \right)}{(v_F
		k_F)^2 - \xi_{\bm{k}}^2} \right] \nonumber\\
	& \simeq  -\rho_0 v_F \int_0^{\omega_D} d \xi_{\bm{k}} \frac{\xi_{\bm{k}}}{x \sqrt{\xi_{\bm{k}}^2
			+ \Delta_0^2}} v_F k_F \frac{\cos (k_F x) \sin \left(
		\frac{\xi_{\bm{k}}}{v_F} x \right)}{(v_F k_F)^2 - \xi_{\bm{k}}^2}\nonumber\\
	& \simeq -\rho_0 \frac{\cos
		(k_F x)}{k_Fx} \int_0^{\omega_D} d \xi_{\bm{k}} \frac{\xi_{\bm{k}}\sin \left( \frac{\xi_{\bm{k}}}{v_F}
		x \right)}{ \sqrt{\xi_{\bm{k}}^2 +(\text{Re}\Delta_0)^2- (\text{Im}\Delta_0)^2+2i\text{Re}\Delta_0\text{Im}\Delta_0}}\,.
	\label{correlation}
\end{align}
Since we expand the energy around the Fermi surface, we assume the condition $\omega_D\ll \mu=v_F k_F$ and neglect the first term in the bracket in the second equality of (\ref{correlation}). Here we also replace $(v_F k_F)^2 - \xi_{\bm{k}}^2$ with $(v_F k_F)^2$ due to this approximation. By defining
\begin{equation}
	k:=\frac{\omega_D}{\text{Im}\Delta_0}=\text{cosh}(\frac{U_R}{\rho_0|U|^2}),a:=\frac{\text{Re}\Delta_0}{\text{Im}\Delta_0},
	\label{ka}
\end{equation} 
we rewrite the integral as
\begin{equation}
	C(\bm{x})\simeq-\rho_0 \frac{\text{Im}\Delta_0\cos(k_F x)}{k_Fx} \int_0^{k} dt \frac{t\sin ( \frac{\text{Im}\Delta_0x}{v_F}
		t )}{ \sqrt{t^2 +a^2- 1+2ia}}\,,
	\label{eq:Cx_near_boundary}
\end{equation}
where we change the integration variable from $\xi_{\bm{k}}$ to $t=\frac{\xi_{\bm{k}}}{\text{Im}\Delta_0} $. Here we separate the correlation function into the real and imaginary parts as 
\begin{equation}
	C(\bm{x})\simeq-\rho_0 \frac{\cos(k_F x)}{k_Fx}(F_1(\bm{x})+iF_2(\bm{x})),
\end{equation}
where
\begin{equation}
	F_1(\bm{x})=\text{Im}\Delta_0\text{Re}\left[\int_0^{k} dt \frac{t\sin ( \frac{\text{Im}\Delta_0x}{v_F}
		t )}{ \sqrt{t^2 +a^2- 1+2ia}}\right],F_2(\bm{x})=\text{Im}\Delta_0\text{Im}\left[\int_0^{k} dt \frac{t\sin ( \frac{\text{Im}\Delta_0x}{v_F}
		t )}{ \sqrt{t^2 +a^2- 1+2ia}}\right]\label{F_1F_2}.
\end{equation}
Since this integral is dominated by the region where $t\simeq\pm\sqrt{1-a^2}$, changing the upper bound of the integral will not influence the long-range behavior of the correlation function. Hence, we set $k\to\infty$ here for convenience. We will illustrate this point in the following numerical simulation with a finite upper bound.
When we consider the correlation function on the phase boundary, we have $a=0$. Hence, the function $F_1(x)$ in the real part of the correlation function takes the form of
\begin{equation}
	F_1(\bm{x})=\text{Im}\Delta_0\text{Re}\left[\int_0^{k} dt \frac{t\sin ( \frac{\text{Im}\Delta_0x}{v_F}
		t )}{ \sqrt{t^2 - 1}}\right]\simeq\text{Im}\Delta_0\int_1^{\infty} dt \frac{t\sin ( \frac{\text{Im}\Delta_0x}{v_F}
		t )}{ \sqrt{t^2 - 1}}.
	\label{definition_of_F1}
\end{equation}
To calculate this function, we introduce another function $G_1(x)$ as
\begin{equation}
	G_1 (x) := \text{Im}
	\Delta_0\int_1^{\infty} dt \frac{\cos ( \frac{\text{Im}\Delta_0x}{v_F}
		t )}{ \sqrt{t^2 - 1}}=-\text{Im}\Delta_0\frac{\pi}{2} N_0 \left( \frac{\text{Im} \Delta_0}{v_F} x \right),
\end{equation}
where $N_0$ is the 0-th order Bessel function of the second kind. The relationship between these two functions is 
\begin{equation}
	F_1 (x) \simeq - \frac{v_F}{\text{Im}\Delta_0} G_1' (x)\,.
\end{equation}
Therefore, the real part of the correlation function is
\begin{equation}
	\mathrm{Re}[C(\bm{x})]\simeq\frac{\pi}{2} \rho_0v_F \frac{\cos (k_F x)}{k_Fx} N_0' \left( \frac{\text{Im}
		\Delta_0}{v_F} x \right)\,.
\end{equation}
When we take the limit $x \rightarrow \infty$, we have
\begin{equation}
	\lim_{x \rightarrow \infty} N_0' (x) = \sqrt{\frac{2}{\pi}} \cos \left( x -
	\frac{\pi}{4} \right) \frac{1}{x^{1 / 2}} - \sqrt{\frac{2}{\pi}} \frac{1}{2}
	\sin \left( x - \frac{\pi}{4} \right) \frac{1}{x^{3 / 2}} \simeq 
	\sqrt{\frac{2}{\pi}} \cos \left( x - \frac{\pi}{4} \right) \frac{1}{x^{1 /
			2}}\,.
	\label{Realexact}
\end{equation}
Thus, the real part of the correlation function has the long-range behavior as
\begin{equation}
	\lim_{x \rightarrow \infty} \left[  \frac{\pi}{2} \rho_0 v_F \frac{\cos (k_F x)}{k_Fx}
	N_0' \left( \frac{\text{Im} \Delta_0}{v_F} x \right) \right] \sim \frac{1}{x^{3 / 2}}.
\end{equation}
Then we turn to consider the function $F_2(x)$ as the imaginary part of the correlation function. On the phase boundary, we can rewrite it as
\begin{equation}
	F_2(\bm{x})=\text{Im}\Delta_0\text{Im}\left[\int_0^{k} dt \frac{t\sin ( \frac{\text{Im}\Delta_0x}{v_F}
		t )}{ \sqrt{t^2 - 1}}\right]=-\text{Im}\Delta_0\int_0^{1} dt \frac{t\sin ( \frac{\text{Im}\Delta_0x}{v_F}
		t )}{ \sqrt{t^2 - 1}}
	\label{definition_of_F2}
\end{equation}
Similarly, we can also define another function $G_2(x)$ as
\begin{equation}
	G_2(x)=-\text{Im}\Delta_0\int_0^{1} dt \frac{\cos ( \frac{\text{Im}\Delta_0x}{v_F}
		t )}{ \sqrt{t^2 - 1}}=\frac{\pi}{2} J_0 \left( \frac{\text{Im}\Delta_0}{v_F} x \right),
\end{equation}
where $J_0$ is the 0-th order Bessel function of the first kind. The relationship between these two functions is also given by
\begin{equation}
	F_2 (x) = - v_F G_2' (x). 
\end{equation}
Then the imaginary part of the correlation function is equivalent to
\begin{equation}
	\mathrm{Im}[C(\bm{x})]\simeq\frac{\pi}{2} \rho_0 v_F\frac{\cos (k_F x)}{k_Fx} J_0' \left( \frac{\text{Im}
		\Delta_0}{v_F} x \right)\,.
\end{equation}
Thus, we have a similar long-range behavior for the imaginary part of the correlation function as
\begin{equation}
	\lim_{x \rightarrow \infty} \left[  \frac{\pi}{2} \rho_0 v_F \frac{\cos (k_F x)}{k_Fx}
	J_0' \left( \frac{\text{Im} \Delta_0}{v_F} x \right) \right] \sim \frac{1}{x^{3 / 2}}\,.
	\label{Imexact}
\end{equation}
To summarize, the correlation function takes the form of
\begin{align}
	\lim_{x\rightarrow\infty}{}_L \left\langle\text{BCS} \right| c_{\sigma}^{\dagger}(\bm{x}) c_{\sigma}(\bm{x}) \left| \text{BCS}
	\right\rangle_R&\simeq \lim_{x\rightarrow\infty}\frac{\pi}{2} \rho_0 v_F \frac{\cos (k_F x)}{k_Fx}( N_0' ( \frac{\text{Im}\Delta_0}{v_F} x )+iJ_0'( \frac{\text{Im}\Delta_0}{v_F} x ))\nonumber\\
	&=:(A(l)+iB(l))x^{-3/2}\,,
	\label{correlation_decay}
\end{align}
where
\begin{align}
	A(l)&=\sqrt{\frac{\pi}{2}}\rho_0 \frac{\sqrt{\text{Im}\Delta_0 v_F}}{k_F} \cos (k_F \frac{v_F}{\text{Im}\Delta_0}l)\cos(l-\frac{\pi}{4})\,,\nonumber\\
	B(l)&=\sqrt{\frac{\pi}{2}}\rho_0 \frac{\sqrt{\text{Im}\Delta_0 v_F}}{k_F} \cos (k_F \frac{v_F}{\text{Im}\Delta_0}l)\sin(l-\frac{\pi}{4})\,,
\end{align}
and $l= \frac{\text{Im}\Delta_0}{v_F} x$. The anomalous dimension is defined by $C(\bm{x})\propto x^{-D+2-\eta}$ \cite{Sachdev:2011uj}, where $D$ is the spatial dimension of the system. From Eq. (\ref{correlation_decay}), we can see that the correlation length diverges on the phase boundary and that the anomalous dimension is given by $\eta=1/2$. In addition, we present the numerical plot of the integrals $F_1$ and $F_2$ on the phase boundary in Fig. \ref{fig2} with detailed fitting parameters shown in Table. \ref{fitting_table_2}. In the numerical calculation we take a finite upper bound $k$ given in Eq. (\ref{ka}). These two functions are related to the correlation function as $C(x)=\rho_0 \frac{\cos (k_F x)}{k_Fx}(F_1(x)+iF_2(x))$. 
The fitting results indicate that $F_1(l)$ is proportional to $\sin(l+\frac{\pi}{4})l^{-0.5}$ and $F_2(l)$ is proportional to $\sin(l+\frac{3\pi}{4})l^{-0.5}$. Those results are consistent with our analytical result in Eq. (\ref{correlation_decay}) and indicate that the correlation length diverges on the phase boundary.

\section{Correlation Function Near the Phase Transition}

To calculate the critical exponent of correlation length, we consider the correlation functions near the phase boundary.  The real part of the gap $\Delta_0$ is given by
\begin{equation}
	\text{Re} [\Delta_0] = \omega_D \frac{\text{sinh} \left( \frac{U_R}{\rho_0 |
			U |^2} \right) \cos \left( \frac{U_I}{ \rho_0 | U |^2}
		\right)}{\left( \text{sinh} \left( \frac{U_R}{\rho_0 | U |^2} \right)
		\cos \left( \frac{U_I}{ \rho_0 | U |^2} \right) \right)^2 + \left(
		\text{cosh} \left( \frac{U_R}{\rho_0 | U |^2} \right) \sin \left(
		\frac{U_I}{ \rho_0 | U |^2} \right) \right)^2}\,.\label{Real_gap}
\end{equation}
For convenience, we here consider the shift of the interaction strength by $\delta U_R\in\mathbb{R}$ from a point $U$ on the phase boundary. It can be replaced by an arbitrary amount $\delta U$ along any direction. Up to the first order of $\delta U_R$, we have $\cos \left(
\frac{U_I}{ \rho_0 | U |^2} \right) = \cos \left( \frac{\pi}{2}-\frac{\pi U_R\delta U_R}{|U|^2} \right)=\sin \left(\frac{\pi U_R\delta U_R}{|U|^2} \right) \simeq \frac{\pi U_R\delta U_R}{|U|^2}$. Then the real part is shown to be proportional to $\delta U_R$:
\begin{equation}
	\text{Re} \Delta_0 \simeq \frac{\pi\omega_D U_R}{| U |^2} \frac{\text{sinh}
		\left( \frac{U_R}{\rho_0 | U |^2} \right)}{\text{cosh}^2 \left(
		\frac{U_R}{\rho_0 | U |^2} \right)} \delta U_R \propto \delta U_R\,.
	\label{define_UR_deviation}
\end{equation}
The imaginary part of the gap remains the same as in Eq. (\ref{IDelta_0}) up to the same order of $\delta U_R$: $\text{Im} \Delta_0 =
\frac{\omega_D}{\text{cosh($\frac{U_R}{\rho_0 | U |^2}$)}}$ and   
\begin{equation}
	a=\frac{\pi U_R}{|U|^2}\text{tanh}(\frac{U_R}{\rho_0|U|^2})\delta U_R
	\label{detailed_expression_a}
\end{equation}
from Eq. (\ref{ka}). Then we turn to the correlation function near the phase boundary.

\begin{table}
	\begin{tabular}{|c|c|}
		\hline 
		\multicolumn{2}{|c|}{$F_{1}(l)$}\tabularnewline
		\hline
		\hline 
		Fitting Function & $F_1(l)=\frac{a_{1}\sin(l+a_{2})}{l^{a_{3}}}$\tabularnewline
		\hline 
		$a_{1}$ & $1.246(1.129,1.363)$\tabularnewline
		\hline 
		$a_{2}$ & $0.786(0.783,0.789)$\tabularnewline
		\hline 
		$a_{3}$ & $0.499(0.485,0.513)$\tabularnewline
		\hline 
		$R^{2}$ & $0.9990$\tabularnewline
		\hline 
	\end{tabular}%
	\begin{tabular}{|c|c|}
		\hline 
		\multicolumn{2}{|c|}{$F_{2}(l)$}\tabularnewline
		\hline 
		\hline
		Fitting Function & $F_{2}(l)=\frac{a_{1}\sin(l+a_{2})}{l^{a_{3}}}$\tabularnewline
		\hline 
		$a_{1}$ & $-1.199(-1.190,-1.209)$\tabularnewline
		\hline 
		$a_{2}$ & $-0.7884(-0.7883,-0.7884)$\tabularnewline
		\hline 
		$a_{3}$ & $0.4924(0.4912,0.4936)$\tabularnewline
		\hline 
		$R^{2}$ & $0.9999$\tabularnewline
		\hline 
	\end{tabular}
	
	\caption{The left and right tables show fitting parameters for $F_1(l)$ and $F_2(l)$ on the boundary, respectively. Here $(\rho_{0}U_{R},\rho_{0}U_{I})=(\frac{1}{\pi},\frac{1}{\pi})$. The values in the parentheses show the range of error bars. The parameter $R^2$ represents the confidence of the fitting, which is defined as the ratio of the sum of squares of the regression (SSR) and the total sum of squares (SST).}
	
	\label{fitting_table_2}
\end{table}
\begin{figure}[t]
	\centering \includegraphics[width=0.6\columnwidth]{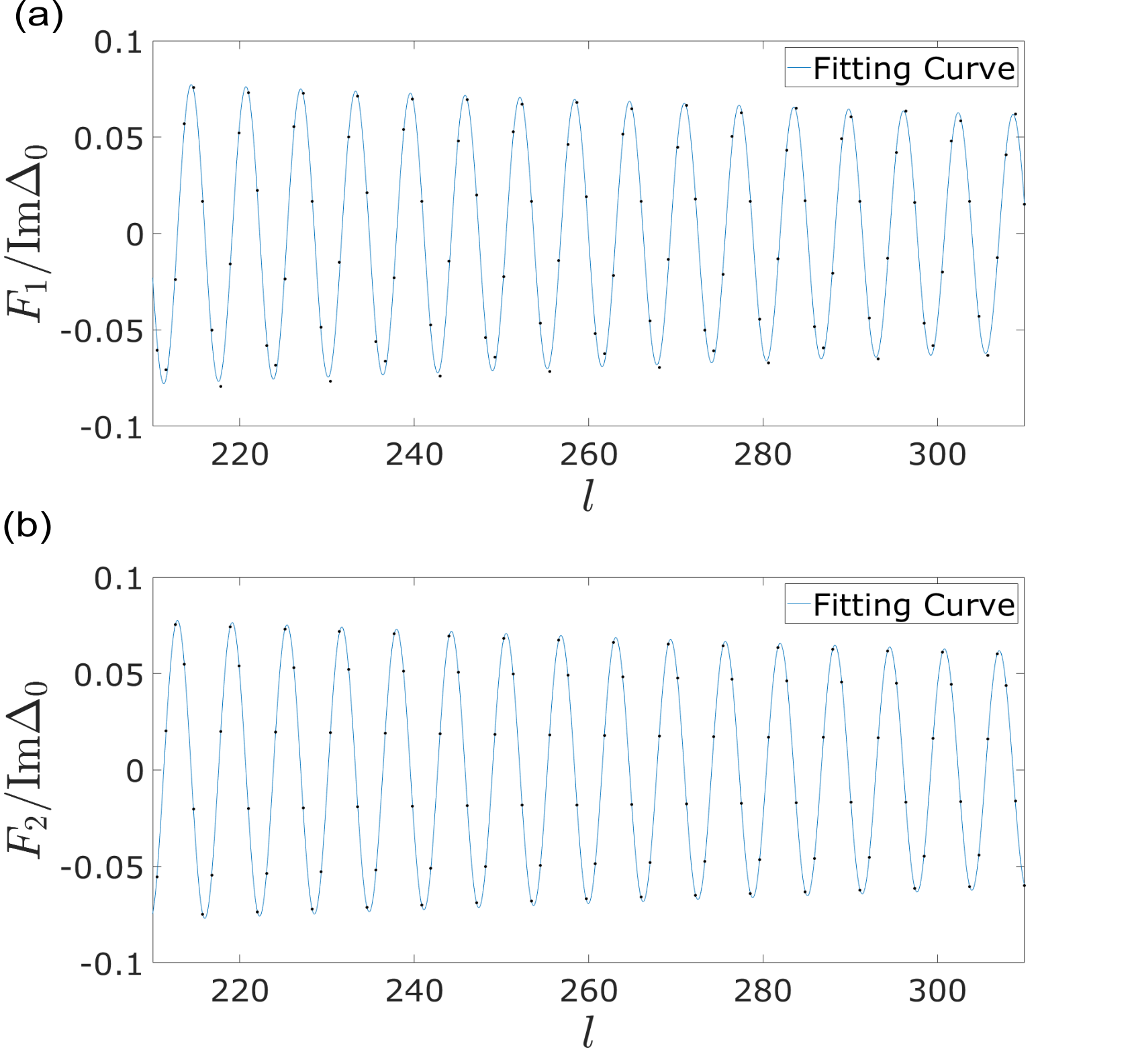}
	\caption{(a) Real part $F_1(l)$ and (b) imaginary part $F_2(l)$ of the correlation function near the boundary (see Eq. (\ref{F_1F_2})). These two figures are near the phase boundary with $(\rho_{0}U_{R},\rho_{0}U_{I})=(\frac{1}{\pi},\frac{1}{\pi})$ with $\rho_{0}\delta U_{R}=1\times10^{-4}\ll1$ and $l=\frac{\text{Im}\Delta_0x}{v_F}.$ The fitting parameters are shown in Table \ref{Fitting_table_1}.}
	\label{fig1}
\end{figure}
\begin{table}
	\begin{tabular}{|c|c|}
		\hline 
		\multicolumn{2}{|c|}{$F_{1}(l)$}\tabularnewline		
		\hline
		\hline 
		Fitting Function & $F_1(l)=a_{1}\sin(a_{2}l-a_{3})e^{-\frac{l}{\xi_{r}}}$\tabularnewline
		\hline 
		$a_{1}$ & $0.1287(0.1153,0.1420)$\tabularnewline
		\hline 
		$a_{2}$ & $1(0.9996,1)$\tabularnewline
		\hline 
		$a_{3}$ & $-0.7843(-0.6798,-0.8887)$\tabularnewline
		\hline 
		$\xi_{r}$ & $420.9(349.6,492.3)$\tabularnewline
		\hline 
		$R^{2}$ & $0.9962$\tabularnewline
		\hline 
	\end{tabular}%
	\begin{tabular}{|c|c|}
		\hline 
		\multicolumn{2}{|c|}{$F_{2}(l)$}\tabularnewline
		\hline
		\hline 
		Fitting Function & $F_2(l)=a_{1}\sin(a_{2}l-a_{3})e^{-\frac{l}{\xi_i}}$\tabularnewline
		\hline 
		$a_{1}$ & $-0.1292(-0.1285,-0.1299)$\tabularnewline
		\hline 
		$a_{2}$ & $1(1,1)$\tabularnewline
		\hline 
		$a_{3}$ & $0.7810(0.7756,0.7864)$\tabularnewline
		\hline 
		$\xi_{i}$ & $417.6(414,421.3)$\tabularnewline
		\hline 
		$R^{2}$ & $0.9937$\tabularnewline
		\hline 
	\end{tabular}
	
	\caption{The left and right tables show fitting parameters for $F_1(l)$ and $F_2(l)$ near the boundary, respectively. Here $(\rho_{0}U_{R},\rho_{0}U_{I})=(\frac{1}{\pi},\frac{1}{\pi})$ with $\rho_{0}\delta U_{R}=1\times10^{-4}\ll1$. The values in the parentheses are the corresponding error bars. The parameter $R^2$ represents the confidence of the fitting, which is defined as the ratio of the sum of squares of the regression (SSR) and the total sum of squares (SST).}
	\label{Fitting_table_1}
\end{table}
The long-range behavior of the correlation functions is governed by the properties of the integral in Eq. (\ref{eq:Cx_near_boundary}). In Fig. \ref{fig1}, we numerically plot $F_1$ and $F_2$ with $(\rho_{0}U_{R},\rho_{0}U_{I})=(\frac{1}{\pi},\frac{1}{\pi})$ and $\rho_{0}\delta U_{R}=1\times10^{-4}\ll1$. $F_1$ is fitted by an oscillating exponential decay  which is shown by the blue curve
in Fig. \ref{fig1}(a): $F_1(l)\propto\sin(l+\frac{\pi}{4})e^{-l/\xi_{r}}$,
where $\xi_{r}=420.9$. $F_2$ behaves similarly as
shown by the blue curve in Fig. \ref{fig1}(b): $F_2(l) \propto\sin(x+\frac{3\pi}{4})e^{-x/\xi_{i}}$
with $\xi_{i}=417.6$. Here $l=\frac{\text{Im}\Delta_0}{v_F}x$ is the same as in the previous section. We can see the behaviors of the real and the
imaginary parts of the correlation function are very close
to each other. The detailed fitting parameters are shown in Table \ref{Fitting_table_1}.

\begin{figure}
	\centering \includegraphics[width=0.6\columnwidth]{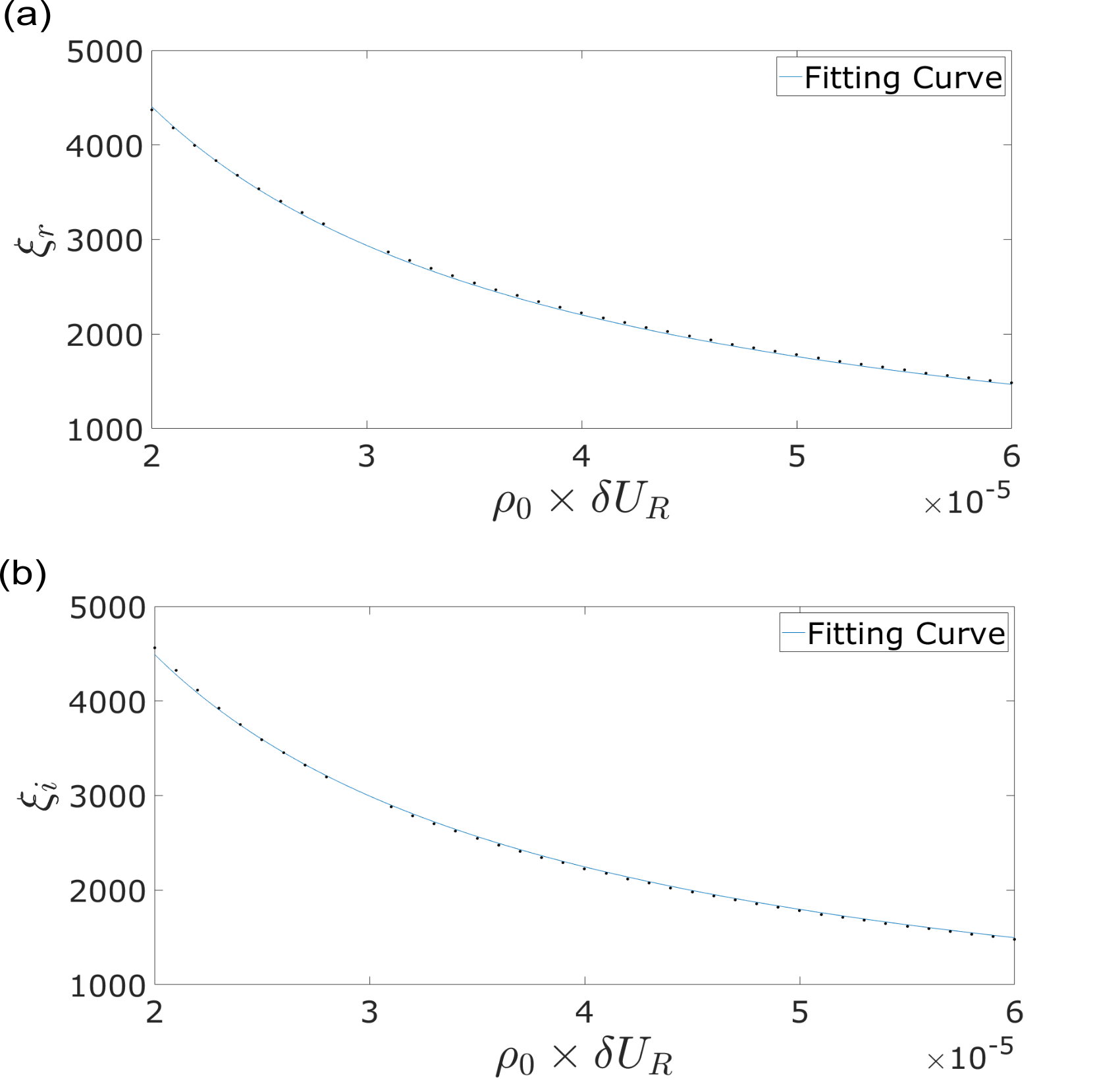}
	\caption{Dependence of the correlation lengths $\xi_r$ and $\xi_i$ on $\rho_{0}\delta U_{R}\ll1$. Here we use the function $f(x)=\frac{a}{x}$ to fit the data with
		$a=0.08815(0.08784,0.08846)$; $0.08985(0.08957,0.09014)$ and $R^{2}=0.9990$;
		$0.9992$, respectively.}
	\label{fig3}
\end{figure}

Furthermore, we numerically calculate the dependence of the correlation lengths on the deviation
$\rho_{0}\delta U_{R}$ from the phase boundary. The correlation lengths of the real part and the imaginary
part are shown separately in Fig. \ref{fig3}. We find that both of the correlation lengths are inversely proportional to the deviation from the phase boundary, i.e. $\xi^{-1}\propto a$. To derive the dependence analytically, we consider the integral $\int_0^{\infty}t\frac{\sin(tx)}{\sqrt{t^2+m^2}}$ with $m\in\mathbb{C}$ and take the limit of $k\to\infty$ in Eq. (\ref{F_1F_2}) since this limiting procedure will not change the long-range behavior. We have
\begin{equation}
	\int_0^{\infty}\frac{t\sin(tx)}{\sqrt{t^2+m^2}}=-\frac{d}{dx}\int_0^{\infty}\frac{\cos(tx)}{\sqrt{t^2+m^2}}=-K_0'(x)\,,
\end{equation}
where $K_{\nu}(x)$ is the $\nu$-th order modified Bessel function of the second kind. By substituting the expression for this special function, we have for $x\rightarrow\infty$
\begin{equation}
	\int_{0}^{\infty}dt\frac{t\sin(xt)}{\sqrt{t^{2}+m^{2}}}\propto \frac{e^{-mx}}{\sqrt{mx}}= \frac{\text{exp}[-\text{Re}(m)x-i\text{Im}(m)x]}{\sqrt{mx}}\,.
\end{equation}
From the definition $C(\bm{x})\propto e^{-x/\xi}$ of the correlation length $\xi$ \cite{Sachdev:2011uj}, we obtain
\begin{equation}
	\xi^{-1}\sim\text{Re}(m)\,.
\end{equation}
From Eq. (\ref{eq:Cx_near_boundary}), the parameter $m$ is given by $m=a+i$. Hence, the relationship between $\xi$ and $a$ is given by
\begin{equation}
	\xi\propto a^{-1},\label{corre_length}
\end{equation}
which agrees with our numerical simulation results in Fig. \ref{fig3}. From the expression of $a$ in Eq. (\ref{detailed_expression_a}), we can see that the correlation length is inversely proportional to the deviation from the phase boundary $\xi^{-1}\propto\delta U_R$, which indicates the critical exponent $\nu=1$ from the definition $\xi\propto(\delta U)^{-\nu}$ \cite{Sachdev:2011uj}. However, on the real axis of the interaction strength $U$, this analysis fails because $\text{Im}\Delta_0=0$ on the whole real axis. From Eq. (\ref{correlation}),  the correlation function for $U_I=0$ is proportional to
\begin{equation}
	C(\bm{x})\propto\int_0^{\infty}\frac{t\sin(xt)}{\sqrt{t^2+s^2}}\,,
\end{equation}
where we redefine $t=\frac{\xi_{\bm{k}}}{v_F}$ and $s=\text{Re}\Delta_0$. On the real axis, the real part of the gap is given by
\begin{equation}
	s=\text{Re}\Delta_0=\frac{\omega_D}{\sinh(\frac{1}{\rho_0U_R})}\,.
\end{equation}
For $U_R\to0$, we have $s=\omega_D\text{exp}(-\frac{1}{\rho_0U_R})\to0$ and thus the correlation length is given by
\begin{equation}
	\xi^{-1}\propto\text{exp}(-\frac{1}{\rho_0 U_R})\,.
\end{equation}
We can find that the correlation length cannot be represented by the polynomial form of $\delta U_R$, which indicates that the critical behavior on the real axis is indeed different from that on the upper half complex plane with $U_I\neq0$. As shown in Sec. \ref{RG}, this difference in the critical behavior can be understood from the RG flow.

\section{Thermodynamic Quantities on the Phase Boundary}

In this section, we calculate critical exponents associated with non-analyticity of thermodynamic quantities on the phase boundary. The condensation energy of the non-Hermitian BCS model is given by \cite{Yamamoto2019}
\begin{equation}
	\Delta E = - \frac{N}{U_R + i U_I} \left( \text{Im} \Delta_0 \right)^2
	- N \int_{- \omega_D}^{\omega_D} d \xi_{\bm{k}} \rho_0 \left( \sqrt{\xi_{\bm{k}}^2 +
		\Delta_0^2} - | \xi_{\bm{k}} | \right)\,.\label{free_energy}
\end{equation}
Note that here we have subtracted the energy of non-interacting fermions from the energy (\ref{free_energy}). The integration is separated into two parts. The second term of the integral is given by
\begin{equation}
	2 \rho_0 \int_0^{\omega_D} d \xi_{\bm{k}} \xi_{\bm{k}} = \rho_0 \omega_D^2\,.
\end{equation}
The first term on the phase boundary is given by
\[ - 2 \int_0^{\omega_D} d \xi_{\bm{k}} \rho_0 \sqrt{\xi_{\bm{k}}^2 + \Delta_0^2} = - 2
\int_0^{\omega_D} d \xi_{\bm{k}} \rho_0 \sqrt{\xi_{\bm{k}}^2 - \left( \text{Im} \Delta_0
	\right)^2}. \]
We first focus on the real part of the energy. Since
\begin{eqnarray}
	\text{Re}\left[- 2 \int_0^{\omega_D} d \xi_{\bm{k}} \rho_0 \sqrt{\xi_{\bm{k}}^2 - \left( \text{Im}
		\Delta_0 \right)^2}\right] & = & - 2 \int_{\text{Im} \Delta_0}^{\omega_D}
	d \xi_{\bm{k}} \rho_0 \sqrt{\xi_{\bm{k}}^2 - \left( \text{Im} \Delta_0 \right)^2}
	\nonumber\\
	& = & (- 2 \rho_0) \left( \text{Im} \Delta_0 \right)^2 \left[ \frac{1}{4}
	\text{sinh} \left( \frac{2 U_R}{\rho_0 | U |^2} \right) - \frac{1}{2}
	\frac{U_R}{\rho_0 | U |^2} \right]\,,
\end{eqnarray}
the real part of the energy is
\begin{eqnarray}
	\text{Re} [\Delta E] & = & - N \frac{U_R}{| U |^2} \left( \text{Im} \Delta_0
	\right)^2 - 2 N \rho_0 \left( \text{Im} \Delta_0 \right)^2 \left[
	\frac{1}{4} \text{sinh} \left( \frac{2 U_R}{\rho_0 | U |^2} \right) -
	\frac{1}{2} \frac{U_R}{\rho_0 | U |^2} \right] + N \rho_0 \omega_D^2
	\nonumber\\
	& = & N \rho_0 \left[ \omega_D^2 - \frac{1}{2} \left( \text{Im} \Delta_0
	\right)^2 \text{sinh} \left( \frac{2 U_R}{\rho_0 | U |^2} \right) \right]
	\nonumber\\
	& = & N \rho_0 \omega_D^2 \left[ 1 - \frac{1}{2} \frac{\text{sinh} \left(
		\frac{2 U_R}{\rho_0 | U |^2} \right)}{\text{cosh}^2 \left( \frac{U_R}{\rho_0
			| U |^2} \right)} \right]\nonumber\\
	& = & N \rho_0 \omega_D^2 \left[ 1-\tanh\left(
	\frac{U_R}{\rho_0 | U |^2} \right)\right]\,.
	\label{free}
\end{eqnarray}

In a similar manner, we calculate the imaginary part of the condensation energy on the phase boundary as
\begin{eqnarray}
	\text{Im} [\Delta E] & = & - \frac{N}{| U |^2} \left( - U_I \right)
	\left( \text{Im} \Delta_0 \right)^2 - 2 N \rho_0 \int_0^{\text{Im} \Delta_0}
	d \xi_{\bm{k}} \sqrt{\xi_{\bm{k}}^2 - \left(\text{Im} \Delta_0\right)^2} \nonumber\\
	& = & \frac{2U_I N}{2 | U |^2} \left( \text{Im} \Delta_0 \right)^2 - 2 N
	\rho_0 \left( \text{Im} \Delta_0 \right)^2 \times \frac{\pi}{4} \nonumber\\
	& = & \frac{N}{2} \rho_0 \left( \text{Im} \Delta_0 \right)^2 \left[
	\frac{2U_I}{\rho_0 | U |^2} - \pi \right] \nonumber\\
	& = & 0 \,.
\end{eqnarray}
Thus, the imaginary part of the condensation energy vanishes on the phase boundary. 

We here relate the number of roots $\chi$ of the partition function with the nonanalyticity of the condensation energy. According to the definition of $\chi$ in the main text, we have
\begin{equation}
	\chi / \beta \simeq\frac{\omega_D}{\pi\text{cosh}(\frac{U_R}{\rho_0|U|^2})}
\end{equation}
in the zero-temperature limit. Using Eq. (\ref{free}), we find
\begin{equation}
	\Delta E=N\rho_0\omega_D^2(1-\sqrt{1-(\frac{\pi\chi}{\beta\omega_D})^2})\,.
\end{equation}
Since $\frac{U_R}{\rho_0|U|^2}\rightarrow\infty$ near $U=0$, the expressions for the condensation energy and the number of roots can be simplified as
\begin{equation}
	\Delta E=2N \rho_0 \omega_D^2 e^{-2\frac{U_R}{\rho_0|U|^2}},\chi/\beta=\frac{2\omega_D}{\pi}e^{-\frac{U_R}{\rho_0|U|^2}}\,.
\end{equation}
Thus, we have
\begin{equation}
	\Delta E=\frac{\pi^2N\rho_0}{2\beta^2}\chi^2\propto(\chi/\beta)^2\,
\end{equation}
near the origin, which means that condensation energy can be related to the number of roots $\chi$. Similarly, the condensation energy on the real axis is given by
\begin{equation}
	\Delta E=-2N \rho_0 \omega_D^2 e^{-\frac{2}{\rho_0 U_R}}.
\end{equation}
Hence, we have $\Delta E=-\frac{\pi^2N\rho_0}{2\beta^2}\chi^2\propto(\chi/\beta)^2$ near the origin. This tells us that the information of condensation energy on the real axis can be read from the number of roots on the complex plane.

Next, we show that the compressibility exhibits critical behavior near the phase boundary. The compressibility is defined by $\kappa=\frac{\partial^2F}{\partial\mu^2}$ where $\xi_{\bm{k}}=\epsilon_{\bm{k}}-\mu$ and $F=-\frac{1}{\beta}\text{log}Z$ is the free energy of the Bogoliubov quasiparticles. We have
\begin{equation}
	\kappa = - \sum_{\bm{k}} \frac{\Delta_{\bm{k}}^2}{(\xi_{\bm{k}}^2 +
		\Delta_{\bm{k}}^2)^{3 / 2}} = - N\int_{- \omega_D}^{\omega_D} \rho_0 d
	\xi_{\bm{k}} \frac{\Delta_0^2}{(\xi_{\bm{k}}^2 + \Delta_0^2)^{3
			/ 2}}\,.
	\label{compressibility}
\end{equation}
Firstly, we consider the compressibility at points on the phase boundary. By substituting the gap in Eqs. (\ref{RDelta_0}) and (\ref{IDelta_0}) into the compressibility (\ref{compressibility}), we can rewrite it as
\begin{equation}
	\kappa= N\int_{- \omega_D}^{\omega_D} \rho_0 d \xi_{\bm{k}}
	\frac{(\text{Im} \Delta_0)^2}{(\xi_{\bm{k}}^2 - (\text{Im}
		\Delta_0)^2)^{3 / 2}}\,.
\end{equation}
This integral diverges since $\text{Im} \Delta_0 < \omega_D$ for all the points on the boundary. Hence, the compressibility exhibits singularity at each point on the boundary. This cannot occur in the Hermitian case since the integral in Eq. (\ref{compressibility}) is finite for a real gap $\Delta_0$. Near the phase boundary with an infinitesimal deviation $\delta U_R$, we obtain
\begin{equation}
	\kappa\simeq N\int_{- \omega_D}^{\omega_D} \rho_0 d \xi_{\bm{k}}
	\frac{(\text{Im} \Delta_0)^2}{(\xi_{\bm{k}}^2 - (\text{Im} \Delta_0)^2
		+ 2 i \text{Re} \Delta_0 \text{Im} \Delta_0)^{3 / 2}} = N\int_{- A}^A \rho_0
	ds \frac{1}{(s^2 - 1 + 2 i \text{Re}
		\Delta_0 / \text{Im} \Delta_0)^{3 / 2}}\,,
\end{equation}
where $A := \cosh \left( \frac{U_R}{\rho_0 | U |^2} \right)$. Since the points near the value $s = 1$ dominantly contribute to the integral, we expand the integral around this point as
\begin{equation}
	\int_{0}^{c} \rho_0 d \delta s \frac{1}{(2 \delta
		s+ 2 i a)^{3 / 2}}=-[\frac{1}{\sqrt{2ia+c}}-\frac{1}{\sqrt{2ia}}],
\end{equation}
where $c$ is a positive constant. Under the limit $a\to0$, the integral is proportional to $a^{-1/2}$. Hence, we obtain the critical exponent $\zeta = 1 / 2$, which is defined as $\kappa \sim (\delta U)^{- \zeta}$ near the phase boundary.

We note that the critical exponents $\eta$ and $\zeta$ are not independent. Here we show the relation between $\eta$ and $\zeta$ for the energy spectrum that can be expanded as $(k-k_E)^{1/n}$ around a gapless point $k=k_E$. When $n$ is an integer, such dispersion relation appears near an $n$-th order exceptional point in non-Hermitian systems \cite{kato1995perturbation}. For those energy spectra, the long-range behavior of the correlation function takes the form as
\begin{equation}
	C(\bm{x})\propto x^{-2+1/n}\,.
\end{equation}
From the definition of the anomalous dimension, we find $\eta=1-\frac{1}{n}$. Similarly, we find that the critical behavior of the compressibility near the phase boundary is 
\begin{equation}
	\kappa\propto\int_{- A}^A \rho_0
	d \xi_{\bm{k}} \frac{1}{(\xi_{\bm{k}}^2 - 1 + 2 i \text{Re}
		\Delta_0 / \text{Im} \Delta_0)^{2-\frac{1}{n}}}\propto(\delta U)^{-1+\frac{1}{n}}\,,
\end{equation}
which indicates that $\zeta=1-1/n$. Thus, we have
\begin{equation}
	\eta=\zeta\,.
\end{equation}

Finally, we discuss the dynamical critical exponent $z$. From the definition of dynamical critical exponent $z$ in Ref. \cite{Sachdev:2011uj}, we here define it as
\begin{equation}
	\text{Re}\Delta_0\propto\xi^{-z}.
\end{equation}
By referencing (\ref{Real_gap}) and (\ref{corre_length}), we obtain
\begin{equation}
	\text{Re}\Delta_0\propto\xi^{-1}\propto\delta U.
\end{equation}
Hence, we have $z=1$.

\section{Pair Correlation Function}

The pair correlation function is defined by
\begin{equation}
	\rho_2 (\boldsymbol{r}_1 \sigma_1, \boldsymbol{r}_2 \sigma_2 ; \boldsymbol{r}_1' \sigma_1', \boldsymbol{r}_2' \sigma_2') := {}_L \langle
	c_{\sigma_1}^{\dagger} (\boldsymbol{r}_1) c_{\sigma_2}^{\dagger} (\boldsymbol{r}_2) c_{\sigma_2'} (\boldsymbol{r}_2') c_{\sigma_1'}(\boldsymbol{r}_1') \rangle_R\,.
\end{equation}
Here we set $\boldsymbol{r}_1 =\boldsymbol{r}_2 =\boldsymbol{R}$ and $\boldsymbol{r}_1'=\boldsymbol{r}_2' = 0$ to consider the correlation between two Cooper pairs. Without loss of generality, we assume $\sigma_1 = \sigma_1'=\uparrow, \sigma_2 = \sigma_2'= \downarrow$. With Wick's theorem, we can simplify the pair correlation function as
\begin{eqnarray}
	\rho_2 (\boldsymbol{R} \uparrow, \boldsymbol{R} \downarrow ; 0 \uparrow, 0
	\downarrow) & = & {}_L \langle c_{\uparrow}^{\dagger} (\boldsymbol{R})
	c_{\downarrow}^{\dagger} (\boldsymbol{R}) c_{\downarrow} (0) c_{\uparrow} (0)
	\rangle_R \nonumber\\
	& = & {}_L \langle c_{\uparrow}^{\dagger} (\boldsymbol{R})
	c_{\downarrow}^{\dagger} (\boldsymbol{R}) \rangle_R {}_L \langle
	c_{\downarrow} (0) c_{\uparrow} (0) \rangle_R + {}_L \langle
	c_{\uparrow}^{\dagger} (\boldsymbol{R}) c_{\uparrow} (0) \rangle_R {}_L
	\langle c_{\downarrow}^{\dagger} (\boldsymbol{R}) c_{\downarrow} (0) \rangle_R\,.
	\label{R0}
\end{eqnarray}
As shown in Eq. (\ref{correlation_decay}), the second term in Eq. (\ref{R0}) decays as $|\bm{R}|^{- 3 / 2}$ on the phase boundary. Thus, in the limit of $|\bm{R}|\to\infty$, the pair correlation function is given by
\begin{eqnarray}
	\lim_{|\bm{R}|\to\infty}\rho_2 (\boldsymbol{R} \uparrow, \boldsymbol{R} \downarrow ; 0 \uparrow, 0
	\downarrow) & = & {}_L \langle c_{\uparrow}^{\dagger} (0)
	c_{\downarrow}^{\dagger} (0) \rangle_R {}_L \langle c_{\downarrow} (0)
	c_{\uparrow} (0) \rangle_R \nonumber\\
	& = & \frac{1}{N^2} \sum_{\boldsymbol{k}_1, \boldsymbol{k}_2} {}_L \langle
	c_{\boldsymbol{k}_1 \uparrow}^{\dagger} c_{-\boldsymbol{k}_1
		\downarrow}^{\dagger} \rangle_R {}_L \langle c_{-\boldsymbol{k}_2 \downarrow}
	c_{\boldsymbol{k}_2 \uparrow}  \rangle_R \nonumber\\
	& = & \left( \frac{\Delta_0}{U} \right)^2 \,,
\end{eqnarray}
where we have used the translational invariance and the definition $\Delta_0 = - \frac{U}{N}
\sum_{\boldsymbol{k}} {}_L \langle c_{-\boldsymbol{k} \downarrow} c_{\boldsymbol{k}\uparrow}  \rangle_R$. We note that the long-distance limit of the pair correlation function does not vanish on the phase boundary:
\begin{equation}
	\lim_{|\bm{R}|\to\infty}\rho_2 (\boldsymbol{R} \uparrow, \boldsymbol{R} \downarrow ; 0 \uparrow, 0
	\downarrow) = - \frac{(\text{Im} \Delta_0)^2}{U^2} \neq 0\,.
\end{equation}
This non-vanishing behavior of the correlation function at the critical point is due to the non-Hermitian nature of the critical phenomenon. In fact, in nonunitary critical phenomena, the correlation function at the critical point can diverge as a function of the distance rather than decay \cite{Fisher:1978vn}.

\end{document}